\documentclass[aps,preprintnumbers]{revtex4}
\usepackage{amsmath,amssymb,mathtools}
\usepackage{graphicx}
\usepackage{hyperref}
\usepackage{subfigure}
\usepackage{comment}
\usepackage{tensor}
\usepackage{cases}
\usepackage{color}
\usepackage{soul}
\usepackage{bm}

\usepackage{diagbox}

\newcommand{\be}{\begin{eqnarray}}
\newcommand{\ee}{\end{eqnarray}}

\newcommand{\simgt}{\lower.5ex\hbox{$\; \buildrel > \over \sim \;$}}
\newcommand{\simlt}{\lower.5ex\hbox{$\; \buildrel < \over \sim \;$}}

\newcommand*{\dif}{\mathop{}\!\mathrm{d}}

\begin{document}

\vspace{2cm}
\title{Gravitational redshift in the void-galaxy cross-correlation function in redshift space}

\vspace{1cm}

\author{Yue Nan and Kazuhiro Yamamoto}
\affiliation{Department of Physical Science, Graduate School of Science,
Hiroshima University, Higashi-Hiroshima, Kagamiyama 1-3-1, 739-8526, Japan}

\begin{abstract}
We construct an analytic model for the void-galaxy cross-correlation function
that enables theoretical predictions of the dipole signal produced
dominantly by the gravitational redshift within voids for the first time.
By extending a theoretical formulation for the redshift-space distortion
of galaxies to include the second order terms of the galaxy peculiar velocity
$\bm v$ and the gravitational potential, we formulate the void-galaxy
cross-correlation function multipoles in the redshift space, the monopole $\xi_0^{(s)}$,
dipole $\xi_1^{(s)}$ and quadrupole $\xi_2^{(s)}$.
We find that the dipole $\xi_1^{(s)}$ is dominated by the gravitational redshift,
which provide a unique opportunity to detect the gravitational potential of voids.
Thus, for the dipole $\xi_1^{(s)}(s)$, the gravitational redshift is crucial.
Although the higher order effect is almost negligible on the monopole $\xi_0^{(s)}$,
it has an influence on the quadrupole $\xi_2^{(s)}$.
The effects from the random velocity of galaxies and the definition
of the void center on the dipole signal are also discussed.
Our model offers a new theoretical probe for the detection of
gravitational redshift with voids and further tests on cosmology and gravity.
\end{abstract}

\maketitle

\def\rpara{{r_{\scriptscriptstyle \|}}}
\def\rperp{{r_{\scriptscriptstyle \bot}}}
\def\kpara{{k_{\scriptscriptstyle \|}}}
\def\kperp{{k_{\scriptscriptstyle \bot}}}
\def\spara{s_{\scriptscriptstyle \|}}
\def\sperp{s_{\scriptscriptstyle \bot}}
\def\qpara{{q_{\scriptscriptstyle \|}}}
\def\qperp{{q_{\scriptscriptstyle \bot}}}
\def\vpara{v_{\scriptscriptstyle \|}}
\def\bfx{{\bm x}}

\section{Introduction}
\label{sec:intro}
The large-scale structure of the Universe is observed in redshift maps
of galaxy redshift surveys such as the Sloan Digital Sky Survey (SDSS).
However, the mapping of galaxies from real space to redshift space
produces statistical anisotropies caused by peculiar velocities relevant
to the gravitational clustering, which is  better known as the redshift-space
distortion (RSD) \cite{Kaiser1987, Hamilton1997}.
Recent galaxy redshift surveys enabled precise measurements of the
RSDs in the galaxy map
\cite{Peacock2001,Guzzo,Yamamoto2008,Beutler,Beutler2,McGill,Rugerri},
which have been
used to constrain important cosmological parameters, such as the
linear growth rate of structure $f$, useful for distinguishing between
various cosmological models ({e.g.,~\cite{Yamamoto2010,Mohammad,DeffM}).
Hence, the RSD in the large-scale structure of galaxies is beneficial
when testing cosmological models, dark matter models, general
relativity and its alternative theories such as modified gravity.
Recently, general relativistic effects and higher order effects
in the redshift map of galaxies have been investigated~\cite{Baccanelli2016a,Bartolo2016,Baccanelli2016b,Bertacca2017,Yoo2014,Yoo2009,Alam1,Alam2,RASERA}.
These works are extensions of precise modeling of galaxy distributions
to include higher order effects of redshift-space distortions and other effects.
The gravitational redshift and the second order Doppler effect
(from the second order peculiar velocity) in the galaxy clusters
are one of such topics~\cite{Wojtak,Zhao,Jimeno,Kaiser,Cai,STYH}.
Measurements of the gravitational redshift with
galaxies associated with clusters have been reported~\cite{Wojtak,Zhao,Jimeno}.
It is worth mentioning that recent research indicated that the gravitational redshift
caused by the local gravitational potential may become a source of bias in the
calibration of cosmological parameters using type Ia supernovae data combined
with other cosmological probes,
provided that we have no prior knowledge about the local gravitational
potential~\cite{Wojtak2015}. This fact
also addresses an importance for the investigation into the} gravitational
redshift on large scales.

The lowest density areas in the large-scale structure
larger than $10h^{-1}$Mpc, i.e., voids, which are one of the characteristic
structures in the Universe, have become a useful tool for
testing cosmological models and gravity theories~\cite{Hamaus,UVP,Mao,Micheletti,VIMOSP,CaiII}.
Precise modeling of voids with redshift-space distortions provides us
with an approach for testing cosmological models.
Accurate models of galaxy distributions of voids in the linear theory of
density perturbations have been developed~\cite{CaiIII,Hamaus,NadaPercival},
and a constraint on the linear growth rate has been obtained
by considering the RSD with voids~\cite{VIMOSP}.
The void-galaxy cross-correlation is utilized to compare galaxy distribution
in redshift space inside and around voids with observations.
The peculiar velocities of the galaxies are essential for void-galaxy cross-correlation
in redshift space; however, these models are usually constructed up to the
leading order of peculiar velocities of galaxies~\cite{NadaPercival}.

In this paper, we investigate possible signatures of the gravitational redshift
and the higher order effect of peculiar velocities in the galaxy
distribution associated with cosmic voids in redshift space, based on the
progress of these previous works.
We focus our analysis on void-galaxy cross-correlation, which
represents the profile of the galaxy distribution of voids.
We develop an analytic theoretical formulation for the void-galaxy cross-correlation
function in redshift space including higher order effects of
the second order terms of the galaxy peculiar velocity and the gravitational potential.
Our model provides us with theoretical predictions for the multipoles of
the void-galaxy cross-correlation function. In particular,
we demonstrate that the dipole component reflects the gravitational redshift of
a void structure for the first time. This aspect has been mostly neglected
in previous works, which
rely on the theory of the first order of the peculiar velocity.

This article is organized as follows. In Sec.~\ref{sec:formu}, we formulate
the redshift space distortion up to the second order of the peculiar
velocity and the gravitational potential. Then, we present an expression
for the void-galaxy cross-correlation function in redshift space.
The multipoles of the void-galaxy cross-correlation function are
defined including the effects of the gravitational potential and
the second order of the peculiar velocity of the spherical coherent
motion of void, as well as random motions.
In Sec.~\ref{sec:resul}, we adopt specific models for the density profile of
a spherically symmetric void which allows us to demonstrate the
multipoles of the void-galaxy cross-correlation function.
In Sec.~\ref{sec:discu}, we discuss the results and the physical reasons for the
behavior of the multipoles of the void-galaxy cross-correlation function.
We also discuss the nontrivial issue of how to choose the center
of a void and its dependence on the results.
Finally, we summarize our results and highlight our
conclusions in Sec.~\ref{sec:concl}. We also briefly discuss prospective applications
of our investigation for a comparison with observations.
In Appendix~\ref{sec:appena}~and~\ref{sec:appenb}, we present complementary derivations of the mathematical formulas in Sec.~II.
In Appendix~\ref{sec:appenc} we demonstrate the profiles of the density contrast, the velocity field, and the gravitational potential that we construct in Sec.~\ref{sec:profile_UVP}.

\section{Formulation}
\label{sec:formu}
We will start with the formulation for the galaxy distribution in redshift space
associated with a void, including the gravitational redshift and the
second order of the peculiar velocity.
We follow the theoretical formulation developed in Ref.~\cite{STYH} by beginning with a brief review of the formulation.
When there is no effect from the gravitational potential and the
peculiar velocity, the relation between the comoving distance
$\chi$ and the redshift $z$ is given by
\begin{eqnarray}
  \chi=\int_0^z{dz'\over H(z')}.
\end{eqnarray}
When there is a shift of the redshift from the gravitational
potential and the peculiar velocity, $\delta z$, the distance
in redshift space can be expressed as
\begin{eqnarray}
  S=\int_0^{z+\delta z}{dz'\over H(z')}\simeq\chi
  +{\delta z\over H(z)}-{H'(z)\over 2H^2(z)}\delta z^2,
\label{eq:Sred}
\end{eqnarray}
where we evaluated the shift in the comoving distance
by $\delta z$ up to the second order.

To include the shift of a photon's energy caused by the
gravitational potential and the Doppler effect of
peculiar velocity, we need to consider the second order terms
of the peculiar velocity. We work within the Newtonian gauge
of cosmological perturbation theory; however, the results
will not depend on this choice because the void relevant
to our problem is of the subhorizon scale.
Denoting the gravitational potential and the peculiar velocity
by $\psi$ and ${\bm v}$, respectively, we may express $\delta z$
up to the order of ${\cal O}(\bm v^2)$ \cite{STYH},
\begin{eqnarray}
  \delta z=(1+z)\left(
    \bm \gamma\cdot {\bm v}+{1\over 2}\bm v^2-\psi \right),
\label{eq:ds}
\end{eqnarray}
where $\bm \gamma$ is the unit vector of the line of sight, and
$\bm \gamma\cdot {\bm v}$ in Eq.~(\ref{eq:ds}) denotes the
usual Doppler effect, while the term $\bm v^2/2$ does the
transverse Doppler effect.
Inserting Eq.~(\ref{eq:ds}) into Eq.~(\ref{eq:Sred}), we may write $S$
up to the order of ${\cal O}({\bm v}^2)$ as
\begin{eqnarray}
  S=\chi+{(1+z)\over H(z)}\left(\bm \gamma\cdot {\bm v}+{1\over 2}\bm v^2+(\bm \gamma\cdot {\bm v})^2
  -\psi \right)
  -{H'(z)\over 2H^2(z)}(1+z)^2(\bm \gamma\cdot {\bm v})^2.
  \label{eq:bigs}
\end{eqnarray}

For convenience, we assign our coordinate system by adopting the plane parallel approximation (distant observer
approximation). Following this assumption,
the coordinates perpendicular to the line-of-sight direction is the
same, and the position of a galaxy in these directions takes the same value
between the redshift space and the real space.
However, the position parallel to the line-of-sight direction shifts,
which is specified by Eq.~(\ref{eq:bigs}).
We adopt a coordinate system with its origin at the center of a void,
and we denote the position of a galaxy as $\vec r$ and $\vec s$, respectively,
in the real space and the redshift space.
We denote the comoving distance of the center of a void by
$\chi_c=\int_0^{z_c}dz'/H(z')$, where $z_c$ is the redshift
of the position of the center in the case, where there is no gravitational redshift effect or the Doppler effect.

As derived in the Appendix~\ref{sec:appena}, when the center of a void and the origin of the
coordinate does not change between the real space and the redshift space,
$\vec r$ and $\vec s$ are related by
\begin{eqnarray}
  \vec s=\vec r+\biggl[
    {\bm \gamma\cdot {\bm v}\over {\cal H}(z)}
    +{1\over 2}{\bm v^2\over {\cal H}(z)}+{(\bm \gamma\cdot {\bm v})^2
      \over {\cal H}(z)}
  -{\psi\over {\cal H}(z)}
  -{H'(z)\over 2{{\cal H}^2(z)}}(\bm \gamma\cdot {\bm v})^2\biggr]\vec \gamma,
  \label{eq:coorred0}
\end{eqnarray}
where we introduced ${\cal H}=aH$. It might be worth mentioning that
the notations $\bm\gamma$ and $\vec \gamma$ are identical for the convenience of expression, as $\bm\gamma=\vec \gamma$.

In the previous equation, we assumed that the center of a void assumes the same position
in real space and redshift space. However,
as we will discuss later, the definition of the center of a void is not a trivial
problem.
We may also consider the case that the center of a void is shifted in the redshift
space, taking the gravitational redshift as $\psi_c/{\cal H}(z_c)$.
In this case, the center of the void is located at the distance
$S_c=\int_0^{z_c}dz'/H(z')-\psi_c/{\cal H}(z_c)$, and the redshift space
and the real space are related by
\begin{eqnarray}
  \vec s=\vec r+\biggl[
    {\bm \gamma\cdot {\bm v}\over {\cal H}(z)}
    +{1\over 2}{\bm v^2\over {\cal H}(z)}+{(\bm \gamma\cdot {\bm v})^2
      \over {\cal H}(z)}
  -{\psi\over {\cal H}(z)}
  -{H'(z)\over 2{{\cal H}^2(z)}}(\bm \gamma\cdot {\bm v})^2+{\psi_c \over {\cal H}(z_c)}\biggr]\vec \gamma.
  \label{eq:coorred}
\end{eqnarray}
Note that Eq.~(\ref{eq:coorred0}) is reproduced by setting
$\psi_c=0$ in Eq.~(\ref{eq:coorred}). In the section below, we present
the formulation with Eq.~(\ref{eq:coorred}).

For convenience, we hereby adopt a convention in our derivation that the notation
$x$ without a top arrow denotes the magnitude $|\vec x|$ for an arbitrary vector
$\vec x$. For instance, we have $s=|\vec s|$ and $r=|\vec r|$.
To formulate the void-galaxy cross-correlation $\xi^{(s)}(s)$ in redshift space,
we need to use the conservation property between redshift space and real space as follows:
\begin{eqnarray}
  1+\xi^{(s)}(s)=\big(1+\xi(r)\big){\rm det}
  \left|{\partial \vec r \over \partial \vec s}\right|,
 \label{eq:conserve}
\end{eqnarray}
where the superscript $(s)$ reminds us that $\xi^{(s)}(s)$ is a quantity
in redshift space. However, like all
the quantities we will introduce in the following sections, this superscript
can be omitted if we remember that
these quantities as a function of redshift space separation $s$, are defined in redshift space.

To obtain $\xi^{(s)}(s)$ in Eq.~(\ref{eq:conserve}), we need to express
the physical space quantity $\xi(r)$ and the transformation
determinant ${\rm det}\left|{\partial \vec r / \partial \vec s}\right|$
using redshift-space quantities related to the separation $s$ within voids.
We start by examining the terms in Eq.~(\ref{eq:coorred}).
To calculate the $\bm \gamma\cdot {\bm v}$ terms in the expression,
we assume that the peculiar velocities of the galaxies associated with
the void yield to the cosmological continuity equation as \cite{Hamaus}
\begin{eqnarray}
  {\bm v}=-{1\over 3}f(z){\cal H}(z)\Delta(r) \vec r,
  \label{eq:pec}
\end{eqnarray}
where the structure linear growth rate $f$ and the average density contrast $\Delta(r)$
of the void within the radius $r$ are involved.
In Ref.~\cite{UVP}, the authors tested the velocity profile in/around the voids in linear theory
as Eq.~(\ref{eq:pec}) against numerical evaluation from simulation, and they found that Eq.~(\ref{eq:pec}),
together with the best-fitted void profile, is consistent with simulation results, where the infall
velocities corresponding to collapsing voids can be reproduced for smaller characteristic void radii.
We will further address this issue subsequently when applying specific void profiles to test our results.

Since the peculiar velocity mainly contributes to RSDs along the line-of-sight direction $\vec \gamma$,
by defining $\rpara \equiv \vec r\cdot \vec \gamma$, we have
\begin{eqnarray}
  &&{\bm v\cdot \bm \gamma\over {\cal H}(z)}
  =-{1\over 3}f(z)\Delta(r)\rpara
  \simeq-{1\over 3}f(z_c)\Delta(r)\rpara
  \equiv\tilde V(z_c,r)\rpara \nonumber,
  \label{ggH}
\end{eqnarray}
where we defined
\begin{eqnarray}
  &&\tilde V(z_c,r)\equiv-{1\over 3}f(z_c)\Delta(r).
  \label{eq:Vdless}
\end{eqnarray}
Here, we adopt the approximation $z \simeq z_c$ since for a certain galaxy
around the void, it is obvious that $z=z(\chi)=z(\chi_c+\rpara)$,
where $\rpara$ is negligible as a tiny quantity compared with
the distance $\chi_c$ from distant observers.
For further calculations, we need investigations into the relations
for quantities $r$ and $s$ between real space and redshift space.
To take the anisotropies related to line-of-sight direction into account,
we define the dimensionless parameter $\mu=\spara/s$, which is the
cosine of the angle between $\vec s$ and $\vec \gamma$ in redshift space.
Then the components that are parallel to $\vec \gamma$ and perpendicular
to $\vec \gamma$ are given as
\begin{eqnarray}
  \spara = s \mu,~~
  \sperp = s \sqrt{1-\mu^2},
  \label{eq:sdecomp1}
\end{eqnarray}
respectively. On the other hand, the relation for parallel and
perpendicular components between redshift space and real space can be written as
\begin{eqnarray}
  \spara=\rpara+\delta \rpara,~~
  \sperp=\rperp,
  \label{eq:sdecomp2}
\end{eqnarray}
where $\delta \rpara$ denotes the shift on $\rpara$ in the redshift coordinate
along the line-of-sight direction caused by the redshift space distortions,
i.e.,
\begin{eqnarray}
  \delta \rpara= {\bm \gamma\cdot {\bm v}\over {\cal H}(z)}
  +{1\over 2}{\bm v^2\over {\cal H}(z)}+{(\bm \gamma\cdot {\bm v})^2
    \over {\cal H}(z)}
-{\psi\over {\cal H}(z)}
-{H'(z)\over 2{{\cal H}^2(z)}}(\bm \gamma\cdot
{\bm v})^2+{\psi_c \over {\cal H}(z_c)}.
\label{eq:drpara}
\end{eqnarray}

Using Eqs.~(\ref{eq:coorred}) and (\ref{ggH})
we can express $\delta \rpara$ as a function of $\tilde V(z_c,r)$, such that
\begin{eqnarray}
  \delta \rpara=    \tilde V(z_c,r)\rpara
  +{1\over 2}{\cal H}(z_c)\tilde V^2(z_c,r)r^2
  +{\cal H}(z_c)\tilde V^2(z_c,r)\rpara^2
  -{\psi(r)\over {\cal H}(z_c)}
  -{H'(z_c)\over 2}\tilde V^2(z_c,r)\rpara^2
  +{\psi_c\over {\cal H}(z_c)}.
\label{eq:srp2}
\end{eqnarray}
In the previous expression of $\delta \rpara$,
except for $\tilde V(z_c,r)\rpara \sim \mathcal{O}(v)$,
all the other terms are of the order $\mathcal{O}(v^2)$,
which is supposed to be the same order of $\psi$ (see also Eq.~(\ref{eq:order_drpara})).
This assumption is broadly used in the following derivations.

Since the leading term for $\delta \rpara$ is $\tilde V(z_c,r)\rpara
\sim \mathcal{O}(v)$, it follows that terms up to the order of
${\cal O}(\delta \rpara^2)$ are
sufficient to contain all $\mathcal{O}(v^2)$ terms; thus, we express
$r$ up to the order of $\mathcal{O}(v^2)$ as
\begin{eqnarray}
  &&r=\sqrt{\rpara^2+\rperp^2}=\sqrt{(\spara-\delta\rpara)^2
  +\rperp^2}=\sqrt{s^2-2\spara\delta \rpara+\delta \rpara^2}
  \nonumber\\
  &&~~
  \simeq s-\mu\delta \rpara+{\delta \rpara^2\over 2s}(1-\mu^2),
  \label{eq:r}
\end{eqnarray}
together with the direct transformation from Eq.~(\ref{eq:sdecomp2}),
\begin{eqnarray}
    \rpara&=&\spara-\delta\rpara.
  \label{eq:rpara}
\end{eqnarray}
Using Eqs.~(\ref{eq:r}) and (\ref{eq:rpara}),
  and keeping terms up to the order of ${\cal O}(v^2)$
  equivalent to ${\cal O}(\tilde V^2)$,
we write
\begin{eqnarray}
  \tilde V(z_c,r)\rpara=\tilde V(z_c,s) \spara-\tilde V(z_c,s)^2\spara-(\tilde
  V(z_c,s) \tilde V'(z_c,s)/s)\spara^3,
\label{eq:Vzr}
\end{eqnarray}
where we use $V'\equiv \partial V / \partial s$ as a convention.
Inserting Eqs.~(\ref{eq:r}) and (\ref{eq:Vzr}) into the expression
for $\delta \rpara$ in Eq.~(\ref{eq:srp2}),
up to the order of $\mathcal{O}(v^2)$, we finally have $\delta \rpara$ as the function of redshift quantities $s$ and $\spara$ as
\begin{eqnarray}
\delta\rpara&=&{\psi_c\over {\cal H}(z_c)}
  +\tilde V(z_c,s)\spara-\tilde V(z_c,s)^2 \spara-(\tilde V(z_c,s)\tilde V'(z_c,s)/s)
  \spara^3
  \nonumber\\
  &&+{1\over 2}{\cal H}(z_c)\tilde V^2(z_c,s)s^2
  +{\cal H}(z_c)\tilde V^2(z_c,s)\spara^2
  -{\psi(s)\over {\cal H}(z_c)}
  -{H'(z_c)\over 2}\tilde V^2(z_c,s)\spara^2.
  \label{eq:rsp}
\end{eqnarray}

Thus, Eq.~(\ref{eq:rpara}) with Eq.~(\ref{eq:rsp}) give the complete
relation between $\rpara$ and $\spara$, together with Eq.~(\ref{eq:sdecomp1}).
Replacing $\spara$ with $\mu$ and $s$ using the relation
${\partial s / \partial \spara}=\mu$,
eventually we can calculate the coordinate transformation
between physical space and redshift space as a function
of the redshift-space quantities $\mu$ and $s$ as follows:
\begin{eqnarray}
  {\rm det}\left|{\partial \vec r\over \partial \vec s}\right|
  &=& \frac{\partial \rpara}{\partial \spara}=\frac{\partial \rpara}{\partial s}\frac{\partial s}{\partial \spara}
  \nonumber \\
  &\simeq
  &1 -\tilde V'(z_c,s)s\mu^2-\tilde V(z_c,s)+\left\{\tilde V(z_c,s)^2\right\} ' s\mu^2
  +\tilde V^2(z_c,s)
+\Bigl\{\tilde V(z_c,s) \tilde V'(z_c,s)/s\Bigr\}'s^3 \mu^4
\nonumber\\
&&
  +3\Bigl\{\tilde V(z_c,s) \tilde V'(z_c,s)/s\Bigr\}s^2 \mu^2
  -{1\over 2}{\cal H}(z_c)\Bigl\{\tilde V^2(z_c,s)s^2\Bigr\}'\mu+
  {\psi'(s)\over {\cal H}(z_c)}\mu
  \nonumber\\
  &&
  +\Bigl\{-{\cal H}(z_c)\tilde V^2(z_c,s)+{H'(z_c)\over 2}\tilde V^2(z_c,s)\Bigr\}'s^2\mu^3
    +2\Bigl\{-{\cal H}(z_c)\tilde V^2(z_c,s)+{H'(z_c)\over 2}\tilde V^2(z_c,s)\Bigr\}s\mu.
    \label{eq:rstrans}
\end{eqnarray}
On the other hand, using the relation Eq.~(\ref{eq:r}),
we can expand $\xi(r)$ around $s$ for a small $\delta s$ while still keeping terms
up to $\mathcal{O}(v^2)$
as
\begin{eqnarray}
  \xi(r)
  &\simeq& \xi(s)  -\xi'(s)\mu\biggl({\psi_c\over {\cal H}(z_c)}
  +\tilde V(z_c,s)s\mu-\tilde V^2(z_c,s)s\mu
  -{\tilde V(z_c,s)\tilde V'(z_c,s)}s^2\mu^3
  +{1\over 2}{\cal H}(z_c)\tilde V^2(z_c,s)s^2
    \nonumber\\
    &&
  +{\cal H}(z_c)\tilde V^2(z_c,s)s^2\mu^2
    -{\psi(s)\over {\cal H}(z_c)}
  -{H'(z_c)\over 2}\tilde V^2(z_c,s)s^2\mu^2\biggr)
  +{\xi'(s)s\over 2}(1-\mu^2)\mu^2 \tilde V(z_c,s)^2
  +{1\over 2}\xi''(s)\mu^2\bigl(\tilde V(z_c,s)s\mu\bigr)^2.
    \nonumber\\
    \label{eq:xir}
\end{eqnarray}

Having calculated $\xi(r)$ with respect to $s$ and $\mu$ in Eq.~(\ref{eq:xir}) and the determinant for coordinate transformation
in Eq.~(\ref{eq:rstrans}), we eventually determine the redshift-space correlation function $\xi^{(s)}(s,\mu)$ through
the conservation property Eq.~(\ref{eq:conserve}),  as
\begin{eqnarray}
  \xi^{(s)}(s,\mu)=&& -1+(1+\xi(s))\biggl\{1-\tilde V +\tilde V^2 +
  \biggl[\Big((H'-3{\cal H})\tilde V - {\cal H}\tilde V's\Big)\tilde V s + {\psi'(s)\over
  {\cal H}}\biggr]\mu +(5\tilde V\tilde V'-\tilde V')s\mu^2
   \nonumber\\
   && +(H'-2{\cal H})\tilde V \tilde V's^2\mu^3 +(-\tilde V \tilde V'+\tilde V'^2 s+\tilde V \tilde V''s)s\mu^4 \biggr\}
   +\xi'(s)\biggl\{\Big[-{1 \over 2}{\cal H}\tilde V^2 s^2 +{\psi(s)-\psi_c \over {\cal H}}\Big]\mu
   \nonumber\\
   &&
   +{1 \over 2}(5\tilde V-2)\tilde V s\mu^2 +{1 \over 2}(H'-2{\cal H})\tilde V^2 s^2 \mu^3
   +(-{1\over2}\tilde V +2 \tilde V' s)\tilde V s \mu^4 \biggr\} +{1 \over 2} \tilde V^2 s^2 \mu^4 \xi''(s).
   \label{eq:xissmu}
\end{eqnarray}
For more details of the derivation process from Eq.~(\ref{eq:bigs}) to Eq.~(\ref{eq:xissmu}), please refer to Appendix \ref{sec:appena}.

In Eq.~(\ref{eq:xissmu}) for $\xi^{(s)}(s,\mu)$, the anisotropies in redshift space associated with the line-of-sight direction are characterized by $\mu$; thus, the multipole components with respect to $\mu$ can be defined as
\begin{eqnarray}
    \xi_{\ell}^{(s)}(s)={1\over 2}\int_{-1}^{+1}\xi^{(s)}(s,\mu)P_\ell(\mu) \dif \mu,
    \label{eq:defmp}
\end{eqnarray}
where $P_\ell(\mu)$ are the Legendre polynomials.
The monopole and quadrupole components can be written as
\begin{eqnarray}
  \xi_0^{(s)}(s)&=&-1+(1+\xi(s))\biggl[1-\tilde V + \tilde V^2
  - \frac{\tilde V'}{3}s +\frac{22}{15}\tilde V
  \tilde V's+\frac{1}{5}\tilde V'^2 s^2 +\frac{1}{5}\tilde V\tilde V'' s^2 \biggr]
  \nonumber\\
  &&
  +\xi'(s)\biggl[\frac{1}{15}\tilde V s \bigl(-5 +11\tilde V  + 6
  \tilde V' s \bigr)\biggr]+{\tilde V ^2 s^2 \over 10}\xi''(s),
  \label{eq:monopole}
\\
  \xi_2^{(s)}(s)&=&(1+\xi(s))\biggl[\frac{2}{105}\Big(-7\tilde V' s+29\tilde V \tilde V' s
   +6\tilde V'^2 s^2 +6\tilde V\tilde V''s^2\Big)\biggr]
  \nonumber\\
  &&
  +\xi'(s)\biggl[{1 \over 105}\tilde V s \bigl(-14+29\tilde V+24\tilde V's \bigr)\biggr]
  +\frac{2}{35}\tilde V^2 s^2 \xi''(s)
  \label{eq:quadrupole},
\end{eqnarray}
which is the generalization of the result by Ref.~\cite{NadaPercival}.
After dropping out $\mathcal{O}(v^2)$ terms in Eq. (\ref{eq:quadrupole}), the coefficients of our equation are
different from those of Eq. (20) in Ref.~\cite{NadaPercival} at first sight, but this is due to the difference of
the definition of multipoles by a factor of $2\ell+1$.
We now focus on the dipole component as follows:
\begin{eqnarray}
  \xi_1^{(s)}(s)=(1+\xi(s))\biggl[{H'-3{\cal H}\over 3}\tilde V^2 s
    +{3H'-11{\cal H}\over 15}
  \tilde V\tilde V's^2+{\psi'(s)\over 3{\cal H}}\biggr]
    +\xi'(s)\biggl[
      {\psi(s)-\psi_c\over 3{\cal H}(z_c)}+{3H'-11{\cal H}\over 30}\tilde V^2 s^2\biggr].
    \label{eq:dipole}
\end{eqnarray}

We can see from the expressions related to the multipoles that the dipole includes the gravitational potential
$\psi$ and $\psi_c$, but the monopole and the quadrupole
does not include terms concerning $\psi$ as well as $\psi_c$.
This fact suggests that only the dipole is influenced by the gravitational potential
among the multipoles up to $\ell=2$.
The dipole is also influenced by the second order terms of the velocity;
hence, it is reasonable to decompose the dipole into terms representing the
contribution from the velocity component $\xi_{1v}^{(s)}(s)$
and the gravitational component $\xi_{1\psi}^{(s)}(s)$ as
\begin{eqnarray}
    \xi_1^{(s)}(s) = \xi_{1\psi}^{(s)}(s)+ \xi_{1v}^{(s)}(s),
    \label{eq:xi1gv}
\end{eqnarray}
where we defined
\begin{eqnarray}
&&\xi_{1\psi}^{(s)}(s) \equiv \frac{\psi'(s)}{3\mathcal{H}}(1+\xi(s))+
\frac{\psi(s)-\psi_c}{3\mathcal{H}}\xi'(s),
  \label{eq:xi1_g}
\\
&&\xi_{1v}^{(s)}(s) \equiv (1+\xi(s))\biggl({H'-3{\cal H}\over 3}\tilde V^2 s+{3H'-11{\cal H}\over 15}
  \tilde V\tilde V's^2\biggr) + {3H'-11{\cal H}\over 30}\tilde V^2 s^2 \xi'(s).
    \label{eq:xi1_v}
\end{eqnarray}
Furthermore, for convenience, we divide $\xi_{1\psi}^{(s)}(s)$
into two terms as
\begin{eqnarray}
  \xi_{1\psi}^{(s)}(s)=\xi_{1\psi 0}^{(s)}(s)+\xi_{1\psi 1}^{(s)}(s),
\label{eq:xi1psi0psi1}
\end{eqnarray}
with
\begin{eqnarray}
&&  \xi_{1\psi 0}^{(s)}(s) \equiv \frac{\psi(s)-\psi_c}{3\mathcal{H}}\xi'(s)
\\
&&  \xi_{1\psi 1}^{(s)}(s) \equiv \frac{\psi'(s)}{3\mathcal{H}}(1+\xi(s)).
\end{eqnarray}

In the previous derivation, the velocity field is supposed to be a coherent field which
follows from the linear theory of density perturbations.
Random motions from the nonlinear effects, which causes the Fingers-of-God (FoG)
effect could also be important. We may include the effect of the random velocities
by following the Gaussian streaming model \cite{NadaPercival,CaiIII}.
This model remaps the galaxy distribution
through a Gaussian random process with velocity dispersion $\sigma_v$.
For simplicity, we adopt the following expression to take the random motions of galaxies
with velocity dispersion into account for $\xi^{(s)}(s,\mu)$ of Eq.~(\ref{eq:xissmu})
\begin{eqnarray}
  1+\xi^{\sigma}(s,\mu) = \int\frac{1+\xi^{(s)}(s^{\sigma},\mu^{\sigma})}{\sqrt{2\pi}\sigma_v} \exp \left(
  -\frac{\vpara^2}{2\sigma_v^2} \right)\dif \vpara,
  \label{eq:gausstreaming}
\end{eqnarray}
with
\begin{eqnarray}
  &&
  s^{\sigma} = \sqrt{\sperp^2+(\spara-{\vpara \over a H})^2},
  \nonumber\\
  &&
  \mu^{\sigma} = \frac{\spara^{\sigma}}{s^{\sigma}}=\frac{\spara-\frac{\vpara}{a H}}
     {\sqrt{\sperp^2+(\spara-\frac{\vpara}{a H})^2}},
  \label{eq:transf}
\end{eqnarray}
where the upper index $\sigma$ takes quantities with a random velocity $\vpara$ into account,
and the transformed quantity $\xi^{\sigma}(s,\mu)$ naturally contains the effect from the
random velocity with the velocity dispersion $\sigma_v^2$.

With the aforementioned transformation, we can define
the RSD multipoles with velocity dispersion similar to Eq.~(\ref{eq:defmp}) as
\begin{eqnarray}
    \xi_{\ell}^{\sigma}(s)={1\over 2}\int_{-1}^{+1}\xi^{\sigma}(s,\mu)P_\ell(\mu) \dif \mu.
    \label{eq:defmpsig}
\end{eqnarray}
For simplicity, the superscript $(s)$ in Eqs.~(\ref{eq:gausstreaming}) and~(\ref{eq:defmpsig})
is omitted since it is well understood that our model for void RSD is constructed
in redshift space.

\begin{table}[t]
  \label{tab:para_plot}
\begin{center}
  \begin{tabular}{ccl}
\hline
\hline
~Parameter~ & ~~~~~~~~~~~~~Value~~~~~~~~~~~~~~& ~Remark~ \\
\hline
$z$ & $0.5$ & Overall redshift \\
$b$ & $2$ & Galaxy linear bias\\
$\gamma$ & $0.55$ & Growth index for growth rate\\
$\Omega_m$ & $0.3$ & Matter density parameter\\
$\Delta_c$ & $-0.4$ & Dark matter void central density\\
$\alpha$ & $3$ & Void shape (steepness) parameter \\
\hline\hline
\end{tabular}
\caption{
Values for the parameters in our demonstration for the profile from
Ref.~\cite{VIMOSP} in Sec.\ref{sec:profile_VIMOSP} . These values were used for the profile in
Eq.~(\ref{eq:profile_Hawken}) unless otherwise expressly stated.
}
\end{center}
\end{table}

\section{Analysis on Specific Models for Void}
\label{sec:resul}
In Sec.~\ref{sec:formu}, we have established the theoretical formulation
for the void-galaxy cross-correlation function in redshift
space and its multipoles, including the second order terms of
the peculiar velocity as well as the gravitational potential.
In this section, we demonstrate the behavior of the multipoles
by adopting specific models for the void density profile.

\subsection{A simple exponential model}
\label{sec:profile_VIMOSP}

To this end, we first adopt the simplest model of a spherical void
density profile proposed in Ref.~\cite{VIMOSP}, which assumes
the integrated density contrast of matter in the form
\begin{eqnarray}
  \Delta(r)=\Delta_c e^{-(r/r_v)^\alpha},
  \label{eq:profile_Hawken}
\end{eqnarray}
where $\Delta_c$, $r_v$, and $\alpha$ are the parameters;
$\Delta_c$ specifies the amplitude of the matter density contrast;
$r_v$ is the characteristic radius of the void; and $\alpha$ characterizes the
steepness of the void wall.
Here, recall that the $\xi(r)$ we construct is the void-galaxy cross-correlation,
where galaxies are biased tracers compared to dark matter. Hence, to evaluate the
gravitational potential which is dominated by dark matter distribution, we need to
include the galaxy bias in our model.
In Ref.~\cite{Pollina}, the authors exhibited by simulation that the void-tracer cross-correlation,
which in our case is just the void-galaxy cross-correlation, can be characterized by a linear relation.
Hence, we simply assume a linear galaxy bias $b$, which follows
\begin{eqnarray}
  \xi(r)=b \delta(r),
  \label{eq:bias}
\end{eqnarray}
where $\delta(r)$ denotes the dark matter density contrast.
In Ref.~\cite{Pollina}, for galaxies as tracers, $b \approx 1.8 \sim 2$ is justified;
thus, we adopt $b=2$ for simplicity. $\Delta_c=-0.4$ and $\alpha=3$ are chosen to accord with
the integrated galaxy density contrast proposed and presented in Ref.~\cite{VIMOSP}.
Here $\Delta(r)$ is related to the matter density contrast $\delta(r)$
and the gravitational potential $\psi$ by the relation
\begin{eqnarray}
&&\Delta(r)={3\over r^3}\int_0^r \dif r'r'^2 \delta(r'),
\label{eq:bigdelta}
\\
&&\triangle \psi(r)=4\pi G a^2\bar\rho_{\rm m}(a)\delta(r),
\end{eqnarray}
where $\bar \rho_{\rm m}(a)$ is the background matter density.
Assuming a spatially flat cosmology with a cosmological constant,
we may write $\bar \rho_{\rm m}(a)=3 H_0^2 \Omega_m / (8\pi G a^3)$, where
$\Omega_m$ is the matter density parameter and $H_0$ is the Hubble parameter
at the present epoch. Then, the density contrast and the gravitational
potential for the void are given by:
\begin{eqnarray}
&&\delta(r)={1\over r^2}{\dif \over \dif r}\left({r^3\Delta(r)\over 3}\right)=
\Delta_c \left(1-{\alpha\over 3}\left({r\over r_v}\right)^\alpha\right)e^{-(r/r_v)^\alpha},
\\
&&\psi(r) =-{3\Omega_m\over 2a}H_0^2
\int_r^\infty \dif r'r'{ \Delta(r')\over 3}
=-{H_0^2r_v^2\over 2}{\Omega_m\Delta_c\over \alpha a}\Gamma(2/\alpha,(r/r_v)^\alpha),
\label{eq:psi_r}
\end{eqnarray}
where $\Gamma(z,a)$ is the incomplete gamma function.

\begin{figure}[t]
\begin{minipage}{0.45\hsize}
  \begin{center}
    \includegraphics[width=80mm]{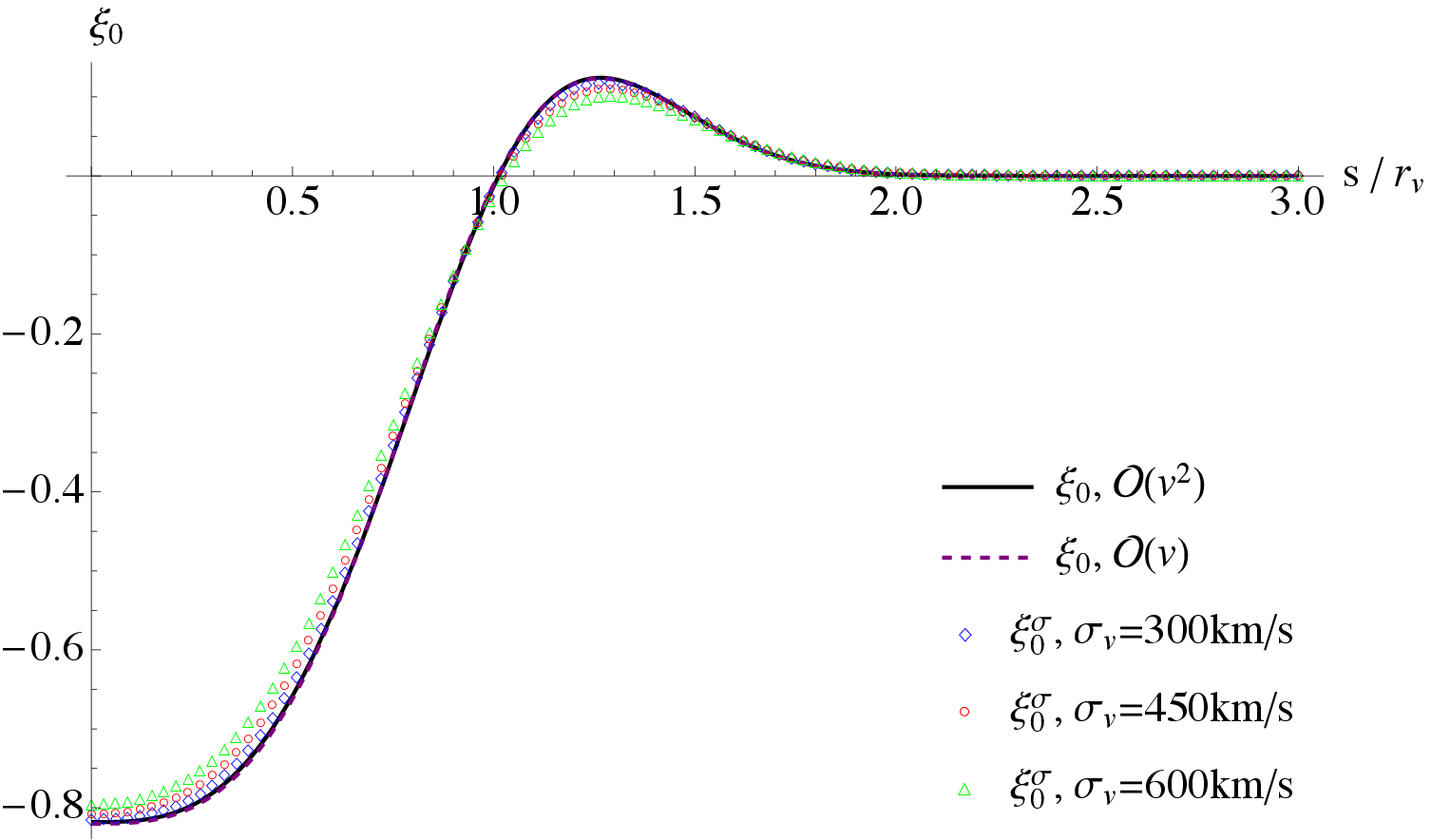}
  \end{center}
      \vspace{-0.cm}
\end{minipage}
\begin{minipage}{0.45\hsize}
  \begin{center}
    \includegraphics[width=80mm]{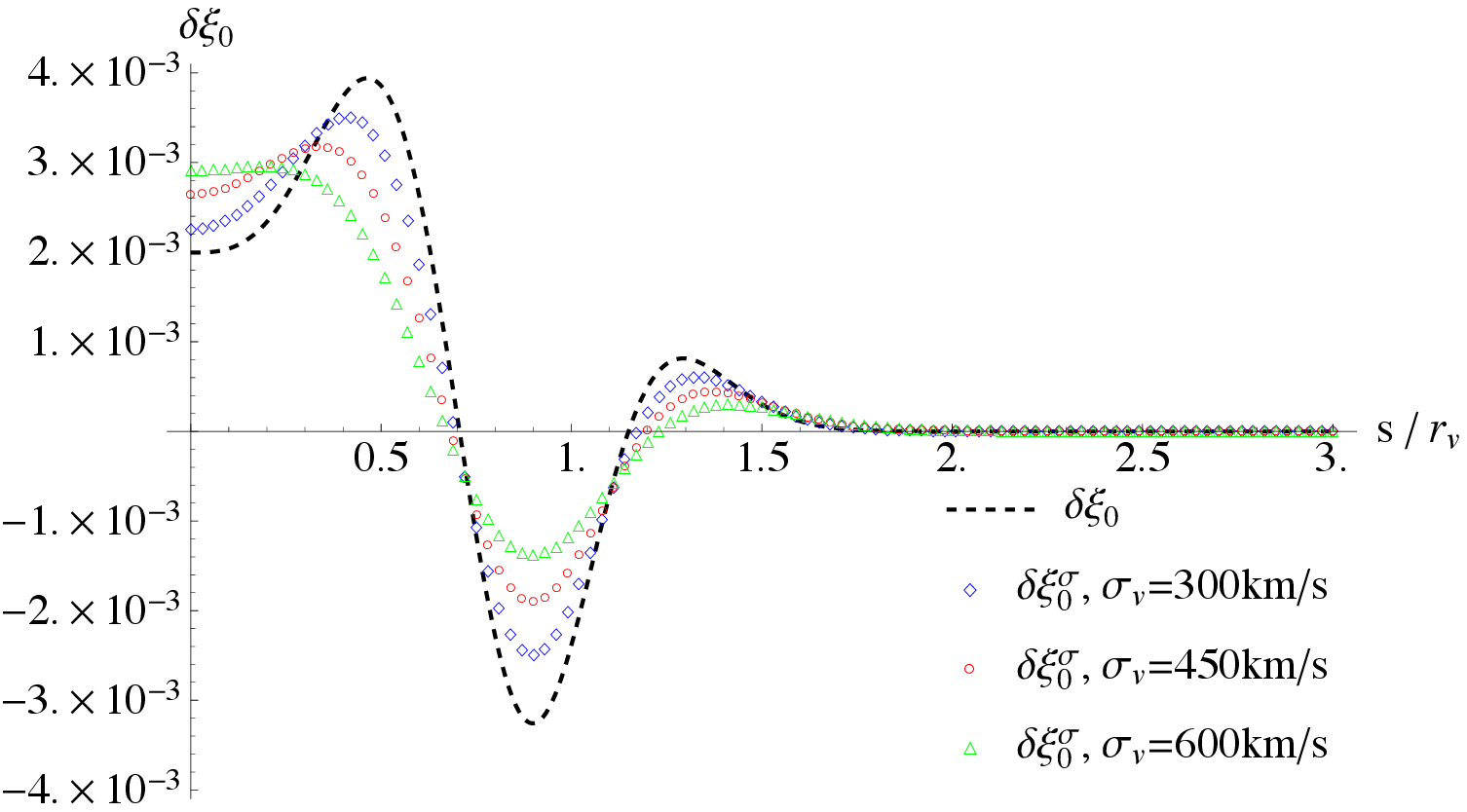}
  \end{center}
      \vspace{-0.cm}
\end{minipage}
\caption{The left panel shows the monopole $\xi_0(=\xi_0^{(s)})$ and $\xi_0^\sigma $ as a function of $s/r_v$.
The solid curve plots $\xi_0$,  Eq.~(\ref{eq:dipole}),
while the symbols show $\xi^{\sigma}_0$, Eq.~(\ref{eq:defmpsig}),
with the different values of the velocity dispersion $\sigma_v$,
whose values are noted in the figure. The dashed curve, which nearly overlaps
with the solid curve, is $\xi_0$ including only the terms up to ${\mathcal O}(v)$.
The right panel shows the difference of the ${\mathcal O}(v^2)$ result and the ${\mathcal O}(v)$ result, defined as
$\delta \xi_0 \equiv \xi_0^{{\mathcal O}(v^2)}- \xi_0^{{\mathcal O}(v)}$.
The symbols in the right panel stand for the
difference of the ${\mathcal O}(v^2)$ result and the ${\mathcal O}(v)$ result
with various values of $\sigma_v$, defined as
$\delta \xi_0^{\sigma} \equiv \xi_0^{\sigma,{\mathcal O}(v^2)}- \xi_0^{\sigma,{\mathcal O}(v)}$.
\label{fig:xi0sig}
}
\vspace{1cm}
\begin{minipage}{0.45\hsize}
  \begin{center}
    \includegraphics[width=80mm]{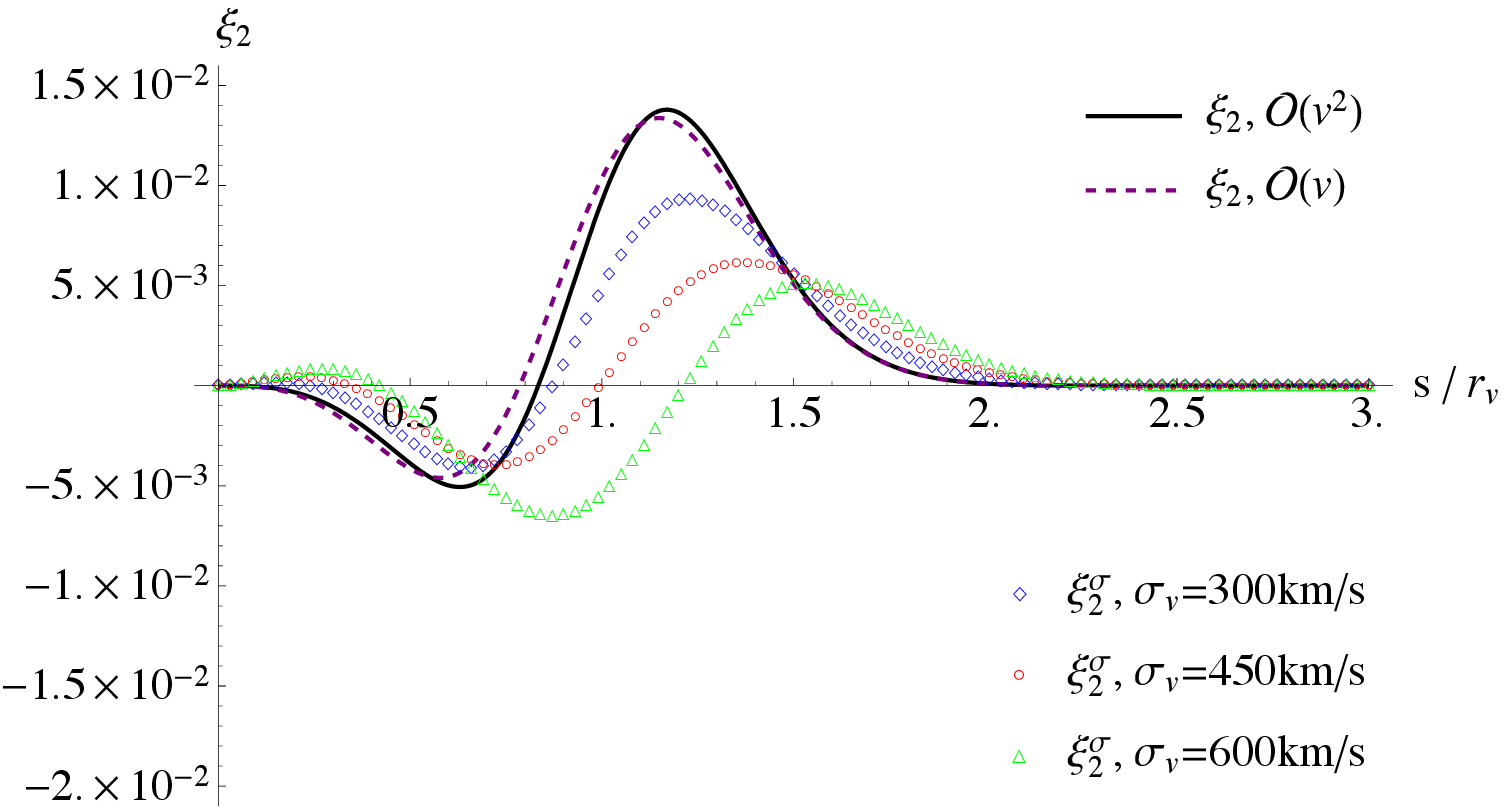}
  \end{center}
      \vspace{-0.cm}
\end{minipage}
\begin{minipage}{0.45\hsize}
  \begin{center}
    \includegraphics[width=80mm]{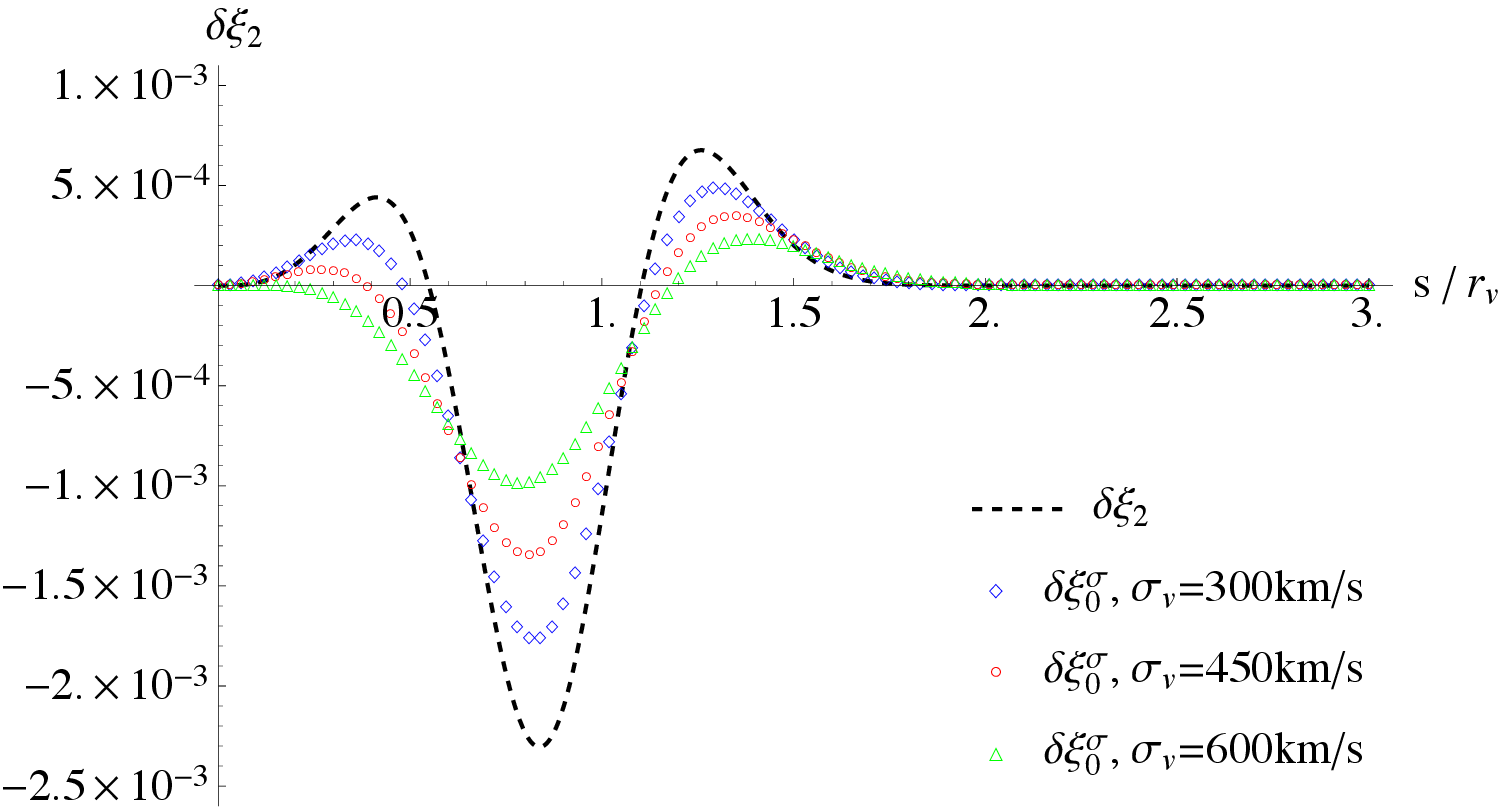}
  \end{center}
      \vspace{-0.cm}
\end{minipage}
\caption{
Same as Figure~\ref{fig:xi0sig} but for the quadrupole
  $\xi_2(=\xi_2^{(s)})$ and $\xi_2^\sigma$ as a function of $s/r_v$ in the left panel.
  The dashed curve is $\xi_2$ including only the terms up to ${\mathcal O}(v)$.
  Again the right panel shows the difference of the ${\mathcal O}(v^2)$ result and the ${\mathcal O}(v)$ result
  as $\delta \xi_2 \equiv \xi_2^{{\mathcal O}(v^2)}- \xi_2^{{\mathcal O}(v)}$.
  Similarly, we have symbols representing the difference of the ${\mathcal O}(v^2)$
  and the ${\mathcal O}(v)$ result
  in the case considering velocity dispersion, where
  $\delta \xi_2^{\sigma} \equiv \xi_2^{\sigma,{\mathcal O}(v^2)}- \xi_2^{\sigma,{\mathcal O}(v)}$.
  }
  \label{fig:xi2sig}
\end{figure}

By solving the continuity equation, the peculiar velocity of the radial direction
can be written (see, e.g., Ref.\cite{VIMOSP}) as
\begin{eqnarray}
&&v(r)=-{\cal H}r
\Delta(r) {f(a)\over3},
\label{PV}
\end{eqnarray}
where $f(a) ={d\ln D_1(a)/d\ln a}$ is the linear growth rate defined by
the logarithmic differentiation with respect to the cosmological scale factor $a$,
which is approximately written as
$f(a)=[\Omega_m(a)]^\gamma$ with $\Omega_m(a)={a^{-3}\Omega_m /({a^{-3}\Omega_m}+1-\Omega_m)}$
and $\gamma=0.55$. We here assume that the coherent peculiar velocity of the galaxies
follows Eq.~(\ref{PV}).
\begin{figure}[ht]
  \begin{center}
  \includegraphics[width=110mm]{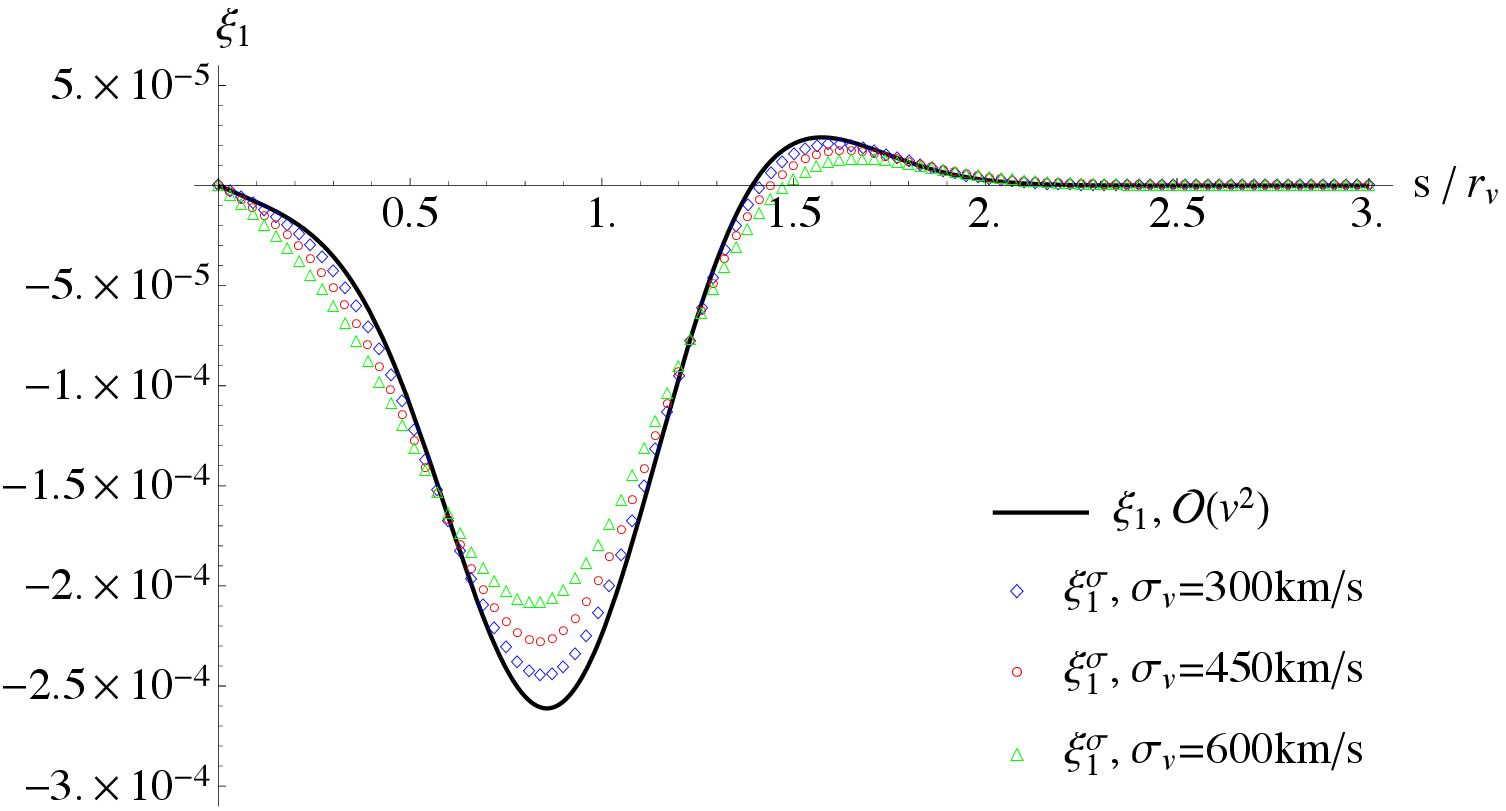}
  \end{center}
  \caption{
    Same as Figure~\ref{fig:xi0sig} but for the dipole component $\xi_1(=\xi_1^{(s)})$
    and $\xi_1^\sigma$ as a function of $s/r_v$.
    The dipole comes from the second order terms of the velocity and the gravitational
    potential.  In this figure, we show the case with the nonzero value of $\psi_c$
    for our model.
  \label{fig:xi1sig}}
\vspace{1cm}
  \begin{center}
  \includegraphics[width=110mm]{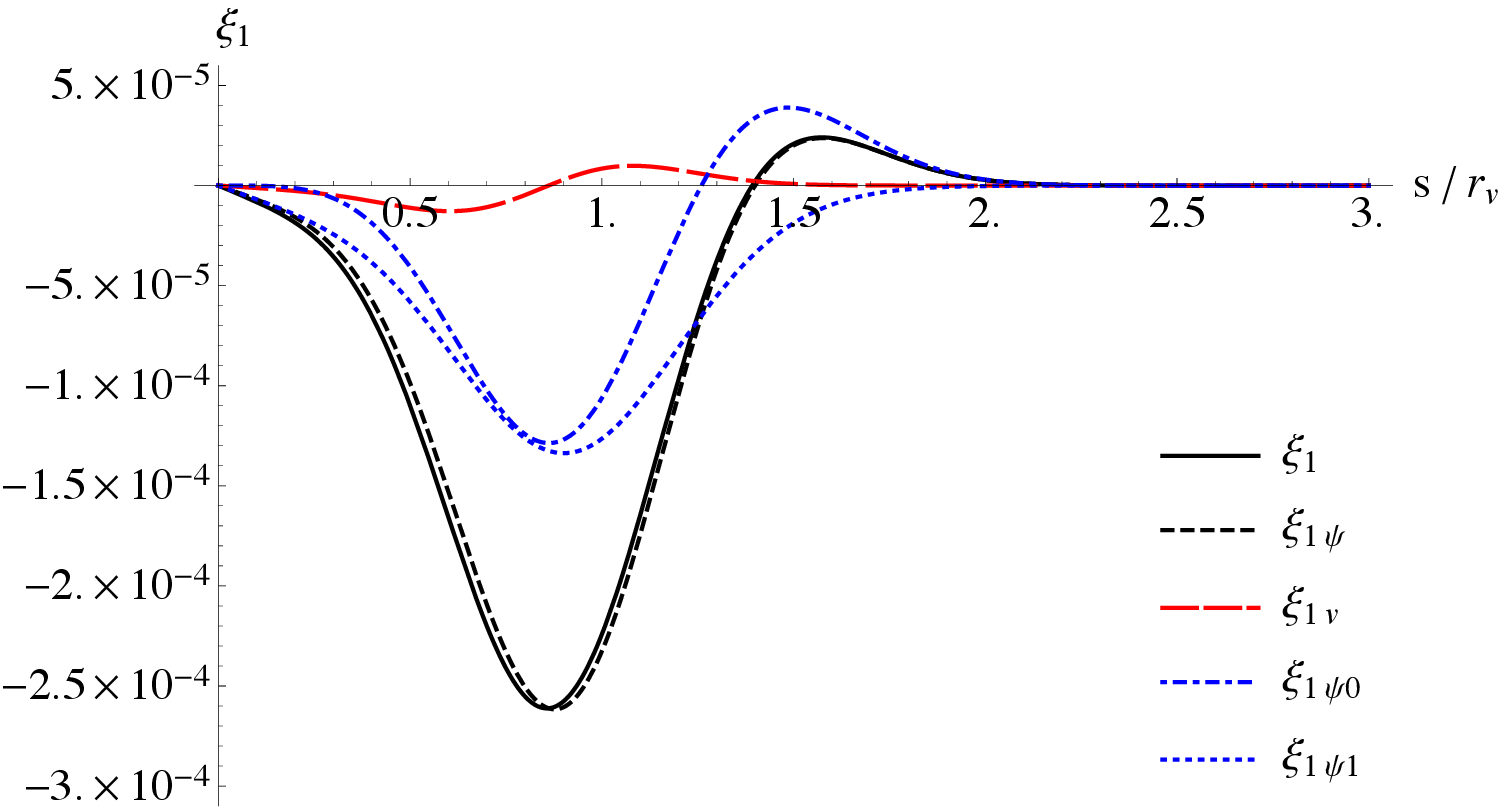}
  \end{center}
 \caption{
   Details of the dipole components, $\xi_{1}(=\xi_{1\psi}^{(s)}+\xi_{1v}^{(s)})$,
   $\xi_{1\psi}^{(s)}(=\xi_{1\psi0}^{(s)}+\xi_{1\psi1}^{(s)})$.
   See Eqs.~(\ref{eq:xi1psi0psi1}) and (\ref{eq:xi1gv}).
  \label{fig:xi1_gvtot_g0g1}}
\end{figure}

Based on the modeling and parametrization above, we are now able to evaluate
the void-galaxy cross-correlation function multipoles.
Figures~\ref{fig:xi0sig} and \ref{fig:xi2sig} plot the
monopole component and the quadrupole component
as functions of the dimensionless characteristic separation $s/{r_v}$,
in correspondence to Eq.~(\ref{eq:monopole}) and Eq.~(\ref{eq:quadrupole}).
In each figure, the solid curve shows the
result including the contribution from the terms up to ${\cal O}(v^2)$,
while the dashed curve shows the result for the terms up to ${\cal O}(v)$.
These two curves overlap in Figure~\ref{fig:xi0sig}, because the deviation is small.
We show the difference for the monopole and quadrupole between $\mathcal{O}(v)$ and $\mathcal{O}(v^2)$
in the right panels of Figures~\ref{fig:xi0sig}~and~\ref{fig:xi2sig}, in which the correction
due to $\mathcal{O}(v^2)$ terms for the monopole is $\sim1\%$, compared with the
$\mathcal{O}(v)$ result.
This suggests that the calculation up to $\mathcal{O}(v)$ is roughly sufficient to predict
the monopole component and that the higher order terms slightly affect the prediction of the quadrupole component.
The symbols in Figures~\ref{fig:xi0sig} and~\ref{fig:xi2sig} represent a plot of the case which
includes the random velocity, Eq.~(\ref{eq:gausstreaming}).
They demonstrate that the effect of the random motion is an important
factor for the quadrupole component.
The monopole and the quadrupole are irrelevant to the gravitational potential
terms in our model; then, we next concentrate mainly on the analysis for the dipole signal.

Figure~\ref{fig:xi1sig} plots the dipole component as a function of $s/r_v$.
The solid curve is the prediction without the random velocity, Eq.~(\ref{eq:dipole}),
while the symbols show the case including the contribution from the random velocity
Eq.~(\ref{eq:gausstreaming}).
For the dipole component, there is no contribution from the order of $\mathcal{O}(v)$.
Figure \ref{fig:xi1_gvtot_g0g1} compares the details of the dipole component.
$\xi_{1v}^{(s)}$ is the contribution from the second order terms of the velocity,
and $\xi_{1\psi0}^{(s)}$ and $\xi_{1\psi1}^{(s)}$ are the contributions from the
gravitational potential and the radial differentiation respectively, which are
defined with Eqs.~(\ref{eq:xi1psi0psi1}) and (\ref{eq:xi1gv}).
This figure shows that $\xi_{1v}^{(s)}$ is a minor contribution
to the dipole and that the dipole is dominated by the contribution
from the gravitational potential.
Furthermore, this figure shows that $\xi_{1\psi}$ arises from the terms
$\psi(s)$ and $\psi'(s)$, which make contributions
to the total result of the dipole at the same level.

\subsection{A universal fitting model}
\label{sec:profile_UVP}
Since the profile of Eq.~(\ref{eq:profile_Hawken}) in Sec.~\ref{sec:profile_VIMOSP}
combined with Eq.~(\ref{eq:pec}) cannot represent the infall velocities for smaller
voids, we consider a best-fit universal void profile for more general cases.
Following the formulation in Ref.~\cite{UVP}, we can write the density contrast
within the void as
\begin{eqnarray}
  \delta(r)=\Delta_c\frac{1-(r/r_s)^{\alpha}}{1+ ({r / r_v})^{\beta}},
  \label{eq:delta_Hamaus}
\end{eqnarray}
where $\alpha$ and $\beta$ are constants; $r_s$ and $r_v$ are some
scale radius and characteristic void radius,
respectively; and $\Delta_c$ is the
central density contrast, similar to Eq.~(\ref{eq:profile_Hawken}).
We define $k\equiv r_s/r_v$ as the ratio of two parameters $r_s$ and $r_v$.
Then the integration within sphere radius $r$ leads to the integrated density contrast
\begin{eqnarray}
 \Delta(r)&=&\frac{3}{r^3}\int^r_0\delta(r')r'^2 \dif r'
 \nonumber\\
 &=&\Delta_c \left[_2F_1\left(1,{3 \over \beta} ; { 3 \over \beta}+1;-(r/r_v)^{\beta}\right)-
 {3 \over \alpha+3}(r/r_v)^{\alpha}k^{-\alpha}{}_2F_1\left(1,{\alpha +3 \over \beta} ;
 {\alpha+3\over\beta}+1;-(r/r_v)^{\beta}\right)\right],
 \label{eq:profile_Hamaus}
\end{eqnarray}
where ${}_2F_1(a,b ; c;x)$ is the hypergeometric function.
In a similar way to Eq.~(\ref{eq:psi_r}), the gravitational potential
characterized by the integrated density contrast Eq.~(\ref{eq:profile_Hamaus})
thus can be written as
\begin{eqnarray}
\psi(r) &=&{\Omega_m\over 2a}H_0^2
\left(\int_0^r \dif r'r'{\Delta(r')}-\int_0^\infty \dif r'r'{\Delta(r')}\right)
\nonumber\\
&=& {\Omega_m H_0^2 \Delta_c(r/r_v)^2  \over 4a} \Bigg\{ 3{}_2F_1\left(1,{2\over\beta} ; {2 \over\beta}+1;
-(r/r_v)^{\beta}\right)-2{}_2F_1\left(1,{3\over\beta} ; {3 \over\beta}+1;-(r/r_v)^{\beta}\right)
\nonumber\\
&&
+6 (r/r_v)^{\alpha}k^{-\alpha}\Bigg[-{1\over\alpha+2}{}_2F_1\left(1,{\alpha+2\over\beta} ;
{\alpha+2\over\beta}+1;-(r/r_v)^{\beta}\right)
\nonumber\\
&&+{1\over\alpha+3}{}_2F_1\left(1,{\alpha+3\over\beta} ; {\alpha+3\over\beta}+1;-(r/r_v)^{\beta}\right)\Bigg]
\Bigg\}-{\Omega_m H_0^2 \over 2a}C_{\psi} ,
\label{eq:psi_r_Hamaus}
\end{eqnarray}
where $C_{\psi}$ is defined as $C_{\psi}\equiv \int_0^\infty \dif r'r'{\Delta(r')}$,
which is parameter dependent and can be determined from the
boundary condition $\psi(\infty)\to 0$. Although analytic calculation for $C_{\psi}$
is difficult, we are able to evaluate it numerically when the parameters are
specified for the void.

We choose three sets of parameters for the different typical size of voids,
the large size void, the medium size void, and the small size void,
which are listed in Table II,
on the basis of the best-fit model in Ref.~\cite{UVP}.
The profiles of the typical voids are presented in Appendix~\ref{sec:appenc} in
Figures~\ref{fig:profile_hamaus} and~\ref{fig:velocity_potential},
with Eqs.~(\ref{eq:delta_Hamaus}),~(\ref{eq:profile_Hamaus}),~and~(\ref{eq:psi_r_Hamaus})
under the deliberate choice of parameters.
The medium size void and the small size void model reproduce the behaviors of
infall velocities presented by simulation in Ref.~\cite{UVP}.

\begin{table}[t]
  \label{tab:para_hamaus}
\begin{center}
  \begin{tabular}{c|ccc|c}
\hline
\hline
\diagbox{Parameters}{Void Size}&~~~Large~~~&~~~Medium~~~&~~~Small~~~&~~~~~Remark~~~~~ \\
\hline
$k$ & $1$ & $0.8725$ &$0.8$&~$r_s/r_v$~\\
$\alpha$ & $2$ & $2.255$ &$2.4$&~$\cdots$~\\
$\beta$ & $8.6$ & $8.76875$ &$7.5$&~$\cdots$~\\
$r_v$ & $30.0$ & $17.6$ &$11.7$&~[$h^{-1}$Mpc]~\\
$\Delta_c$ & $-0.35$ &$-0.43$ &$-0.45$&~$\cdots$~\\
$C_{\psi}$ & $-0.188906$& $-0.0744112$ & $0.242076$ & ~[$h^{-2}$Mpc$^{2}$]~\\
\hline\hline
\end{tabular}
  \caption{The parameters for the large size void, the medium size void, and the
    small size void which accord with the best-fit values in Ref.~\cite{UVP}
    for Eq.~(\ref{eq:pec}) and Eq.~(\ref{eq:psi_r_Hamaus}) in Sec.~\ref{sec:profile_UVP}.
    It is worth mentioning that these three voids represent the typical voids
    of different sizes in Ref.~\cite{UVP}.
    For the medium size void, we choose the parameters to meet the condition
    $\psi_{\rm min}=-\psi(0)$.}
\end{center}
\end{table}

\begin{figure}
\begin{minipage}{0.3\hsize}
  \begin{center}
  \includegraphics[width=\textwidth,height=3.5cm]{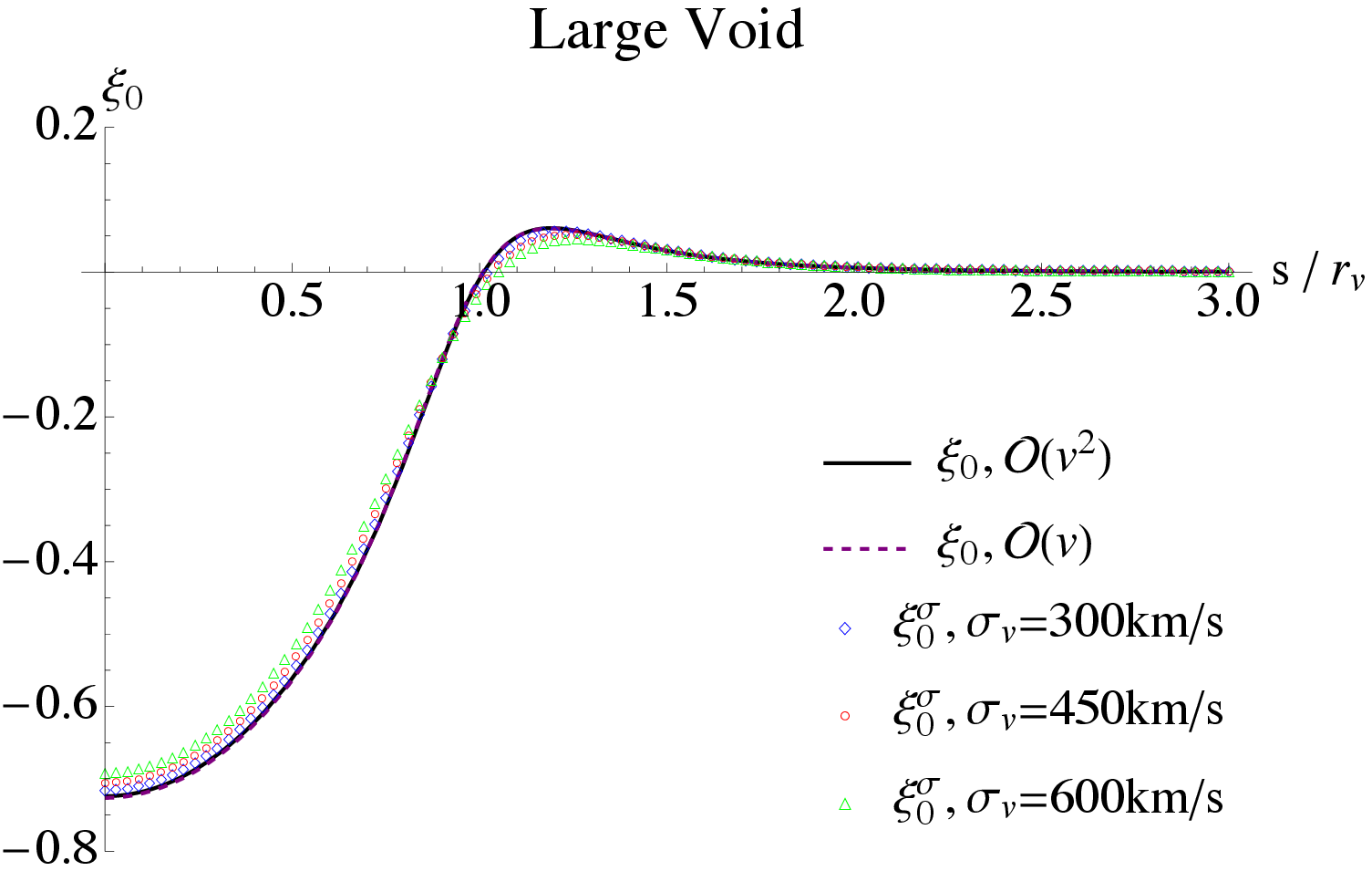}
  \end{center}
   \vspace{-0.cm}
 \end{minipage}
\hspace{0cm}
\begin{minipage}{0.3\hsize}
  \begin{center}
  \includegraphics[width=\textwidth,height=3.5cm]{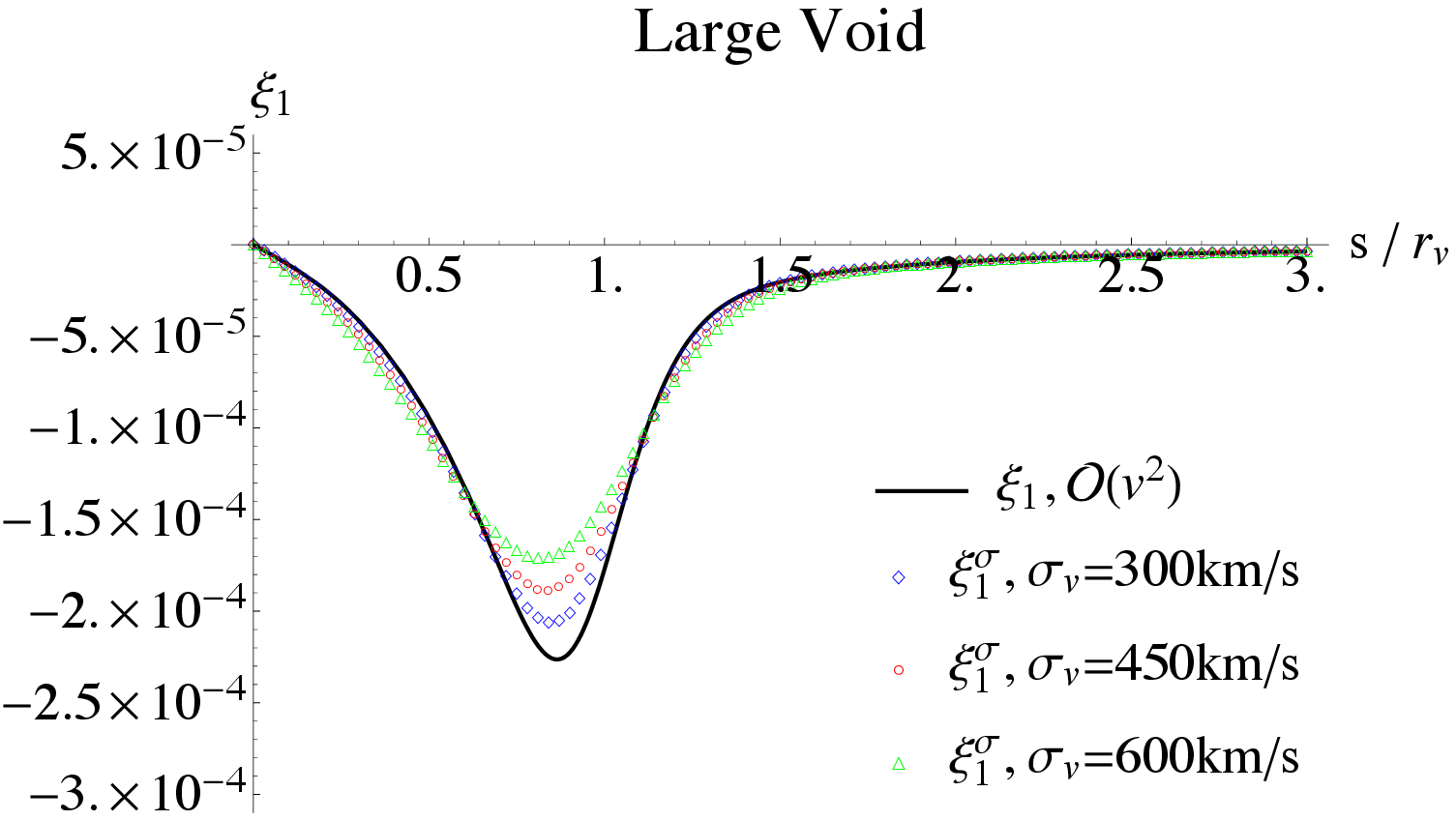}
  \end{center}
   \vspace{-0.cm}
 \end{minipage}
\hspace{0.2cm}
 \begin{minipage}{0.3\hsize}
  \begin{center}
  \includegraphics[width=\textwidth,height=3.5cm]{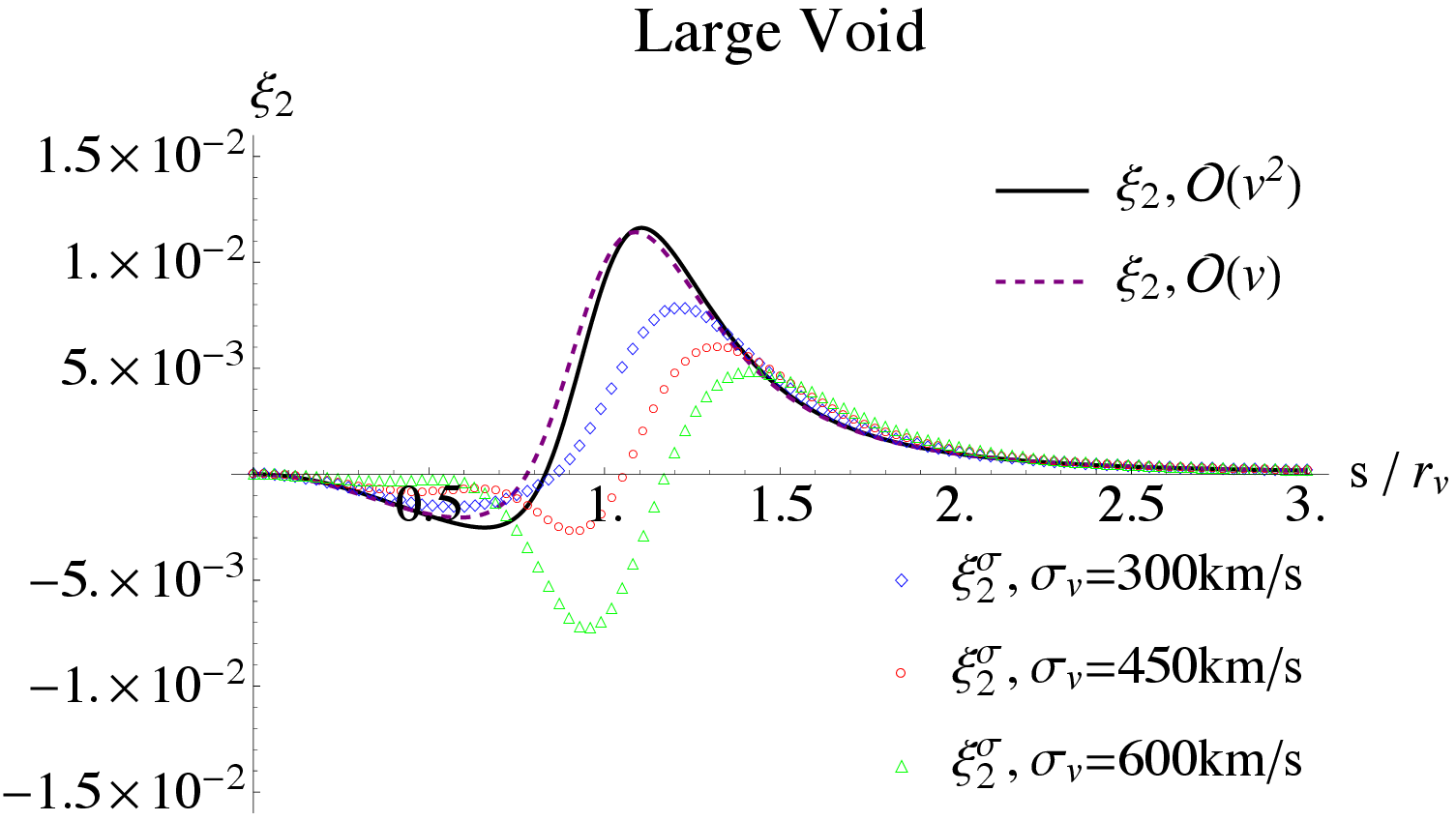}
  \end{center}
   \vspace{-0.cm}
 \end{minipage}
 \begin{minipage}{0.3\hsize}
  \begin{center}
  \includegraphics[width=\textwidth,height=3.5cm]{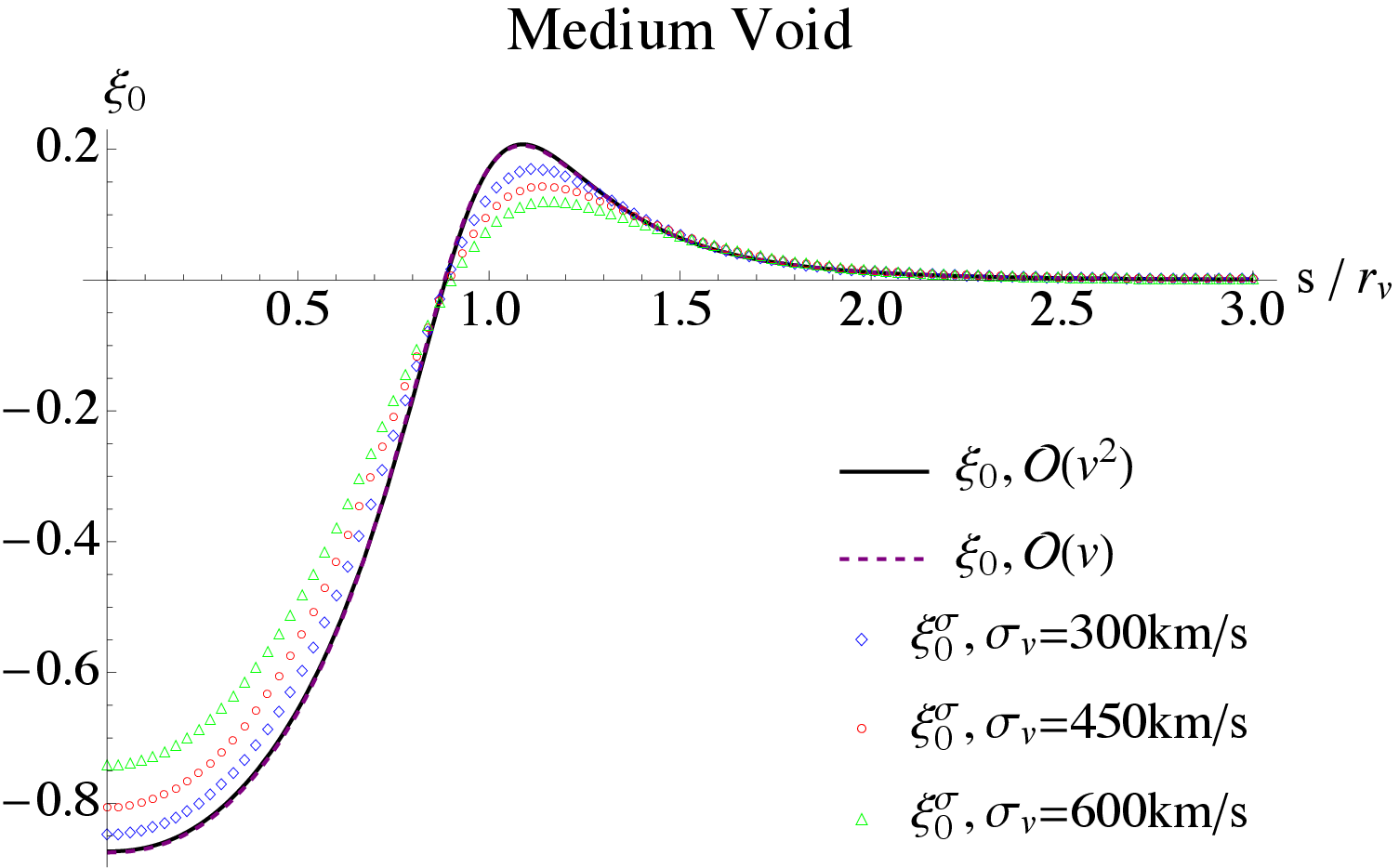}
  \end{center}
   \vspace{-0.cm}
 \end{minipage}
 \hspace{0cm}
 \begin{minipage}{0.3\hsize}
  \begin{center}
  \includegraphics[width=\textwidth,height=3.5cm]{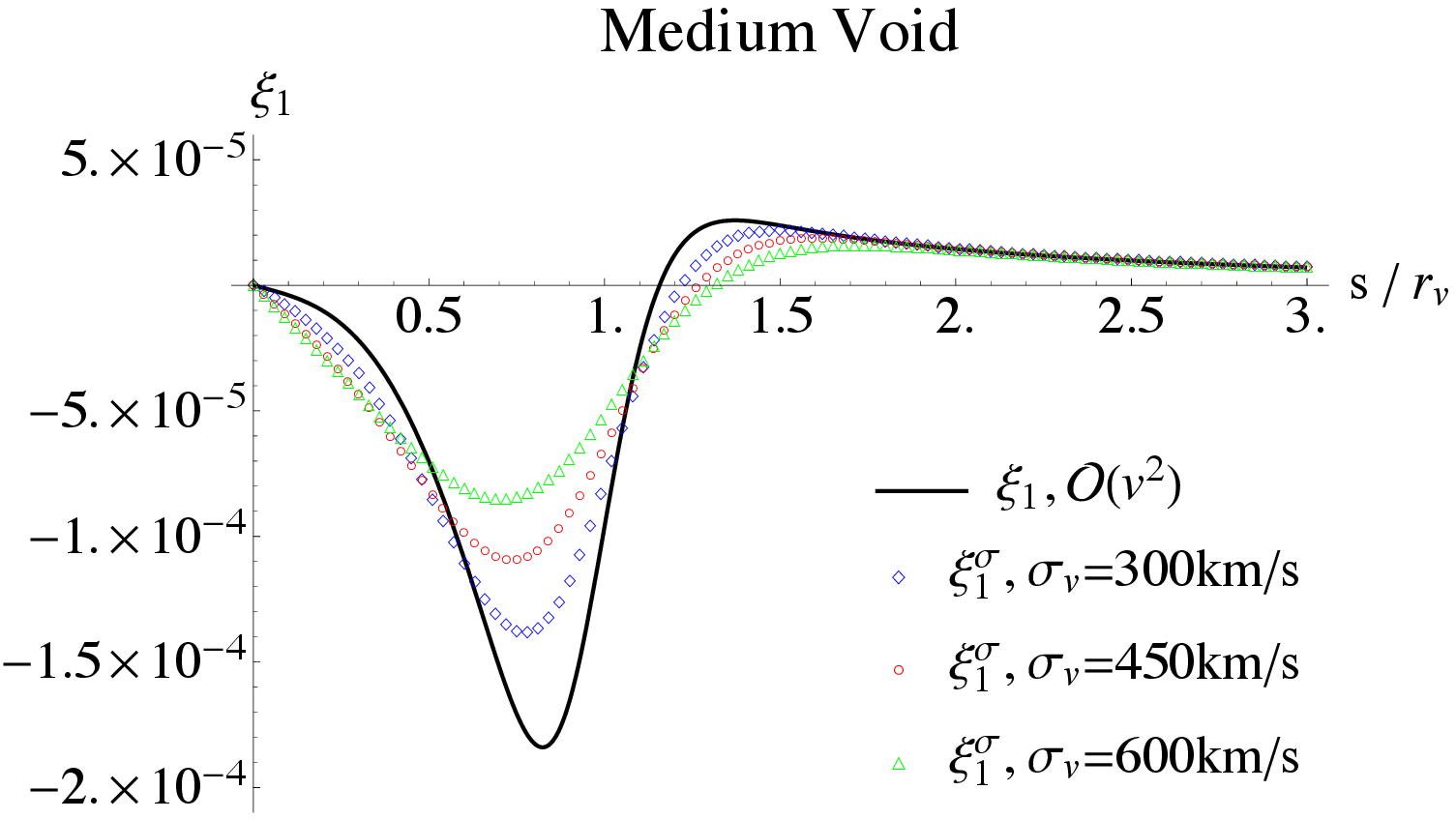}
  \end{center}
   \vspace{-0.cm}
 \end{minipage}
\hspace{0.2cm}
 \begin{minipage}{0.3\hsize}
  \begin{center}
  \includegraphics[width=\textwidth,height=3.5cm]{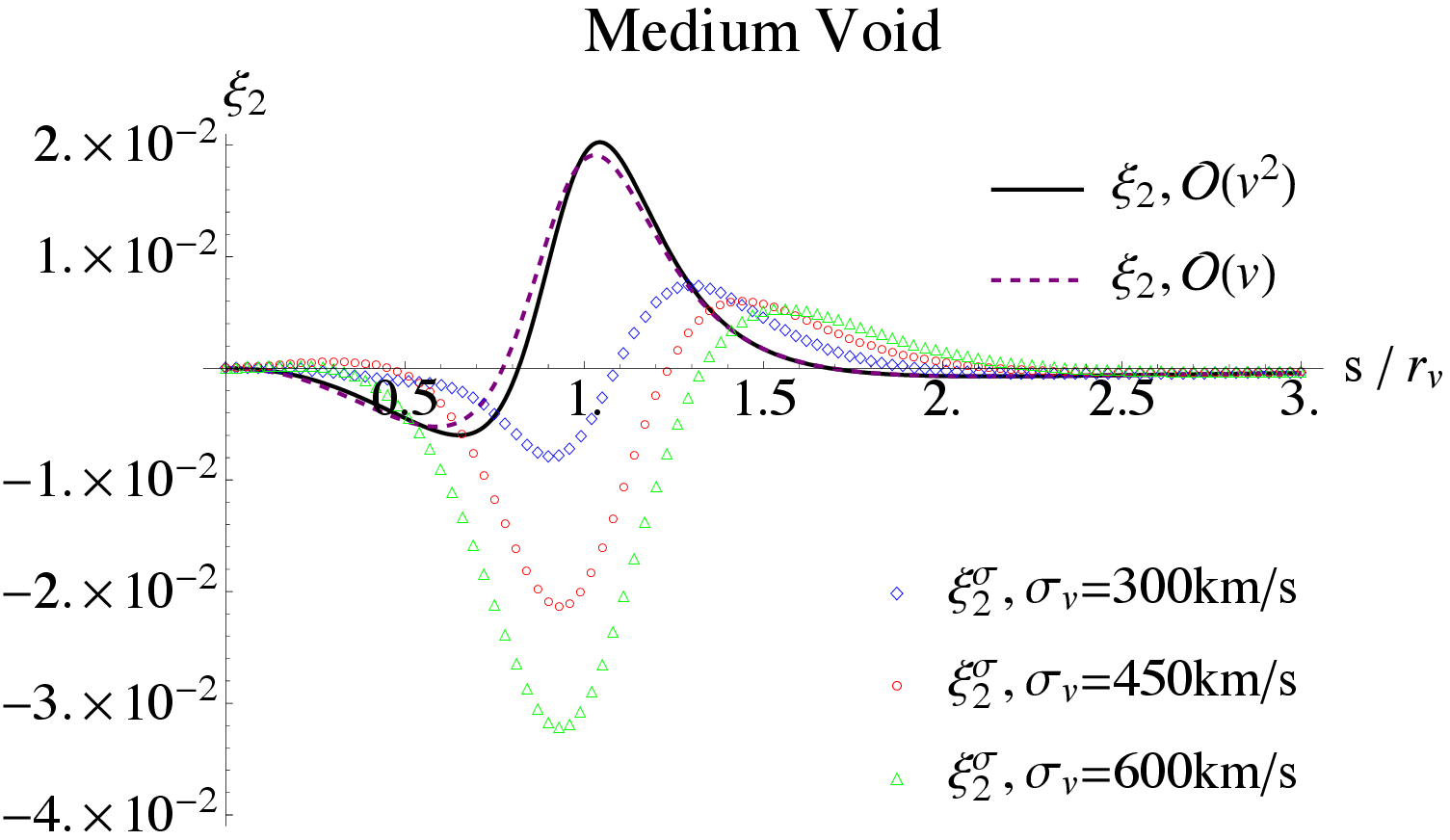}
  \end{center}
   \vspace{-0.cm}
 \end{minipage}
 \begin{minipage}{0.3\hsize}
  \begin{center}
  \includegraphics[width=\textwidth,height=3.5cm]{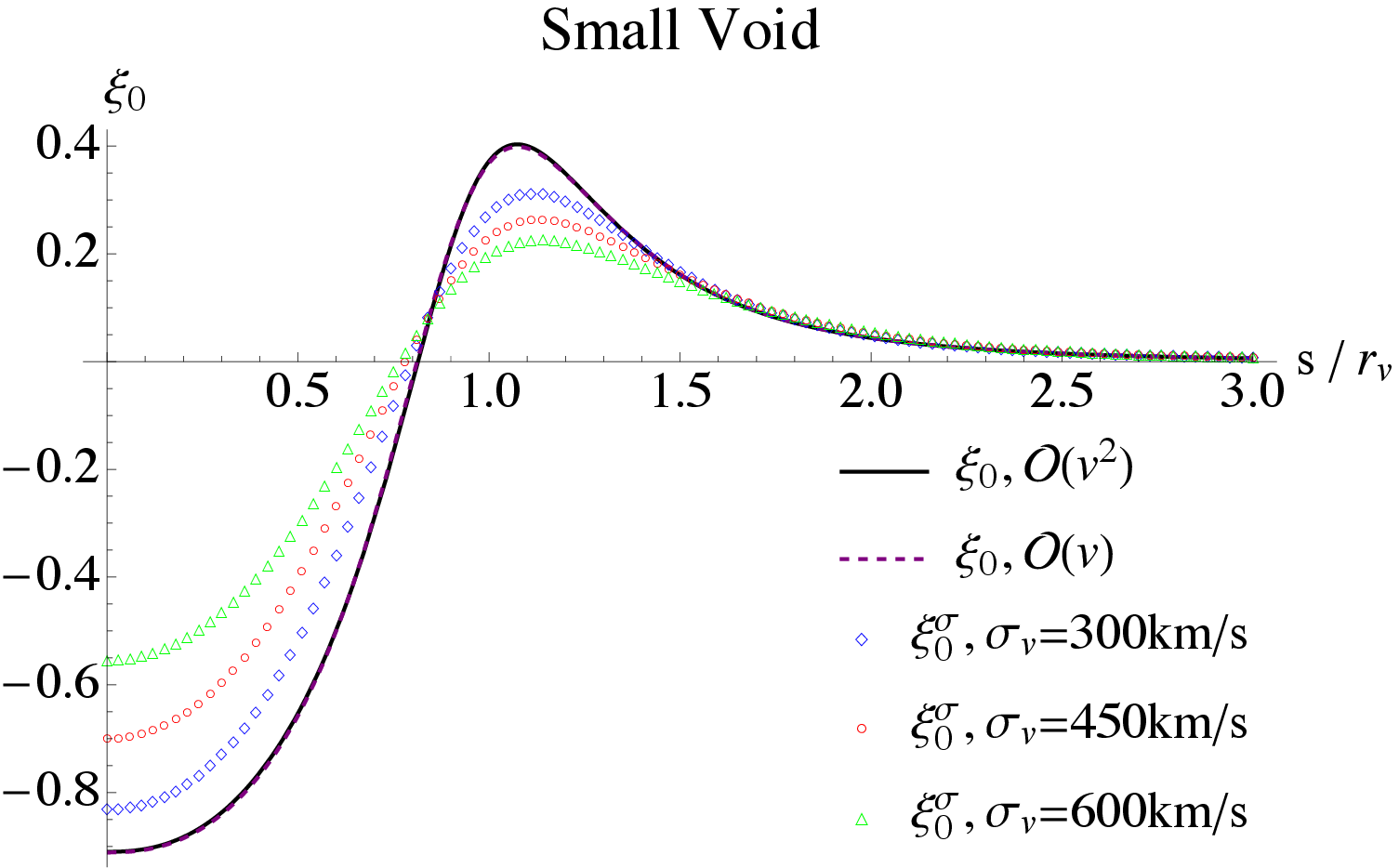}
  \end{center}
   \vspace{-0.cm}
 \end{minipage}
 \hspace{0cm}
 \begin{minipage}{0.3\hsize}
  \begin{center}
  \includegraphics[width=\textwidth,height=3.5cm]{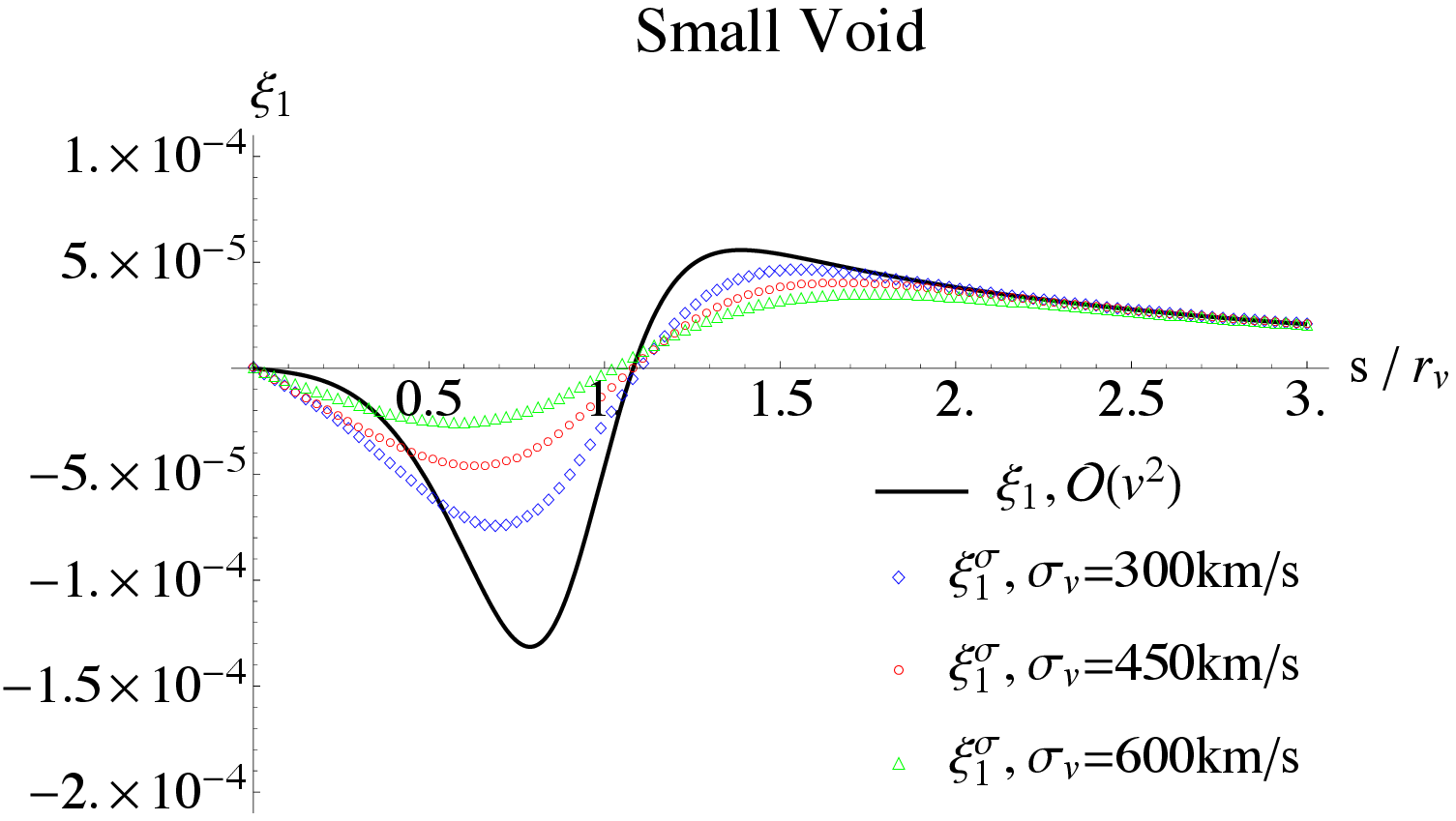}
  \end{center}
   \vspace{-0.cm}
 \end{minipage}
 \hspace{0.2cm}
 \begin{minipage}{0.3\hsize}
  \begin{center}
  \includegraphics[width=\textwidth,height=3.5cm]{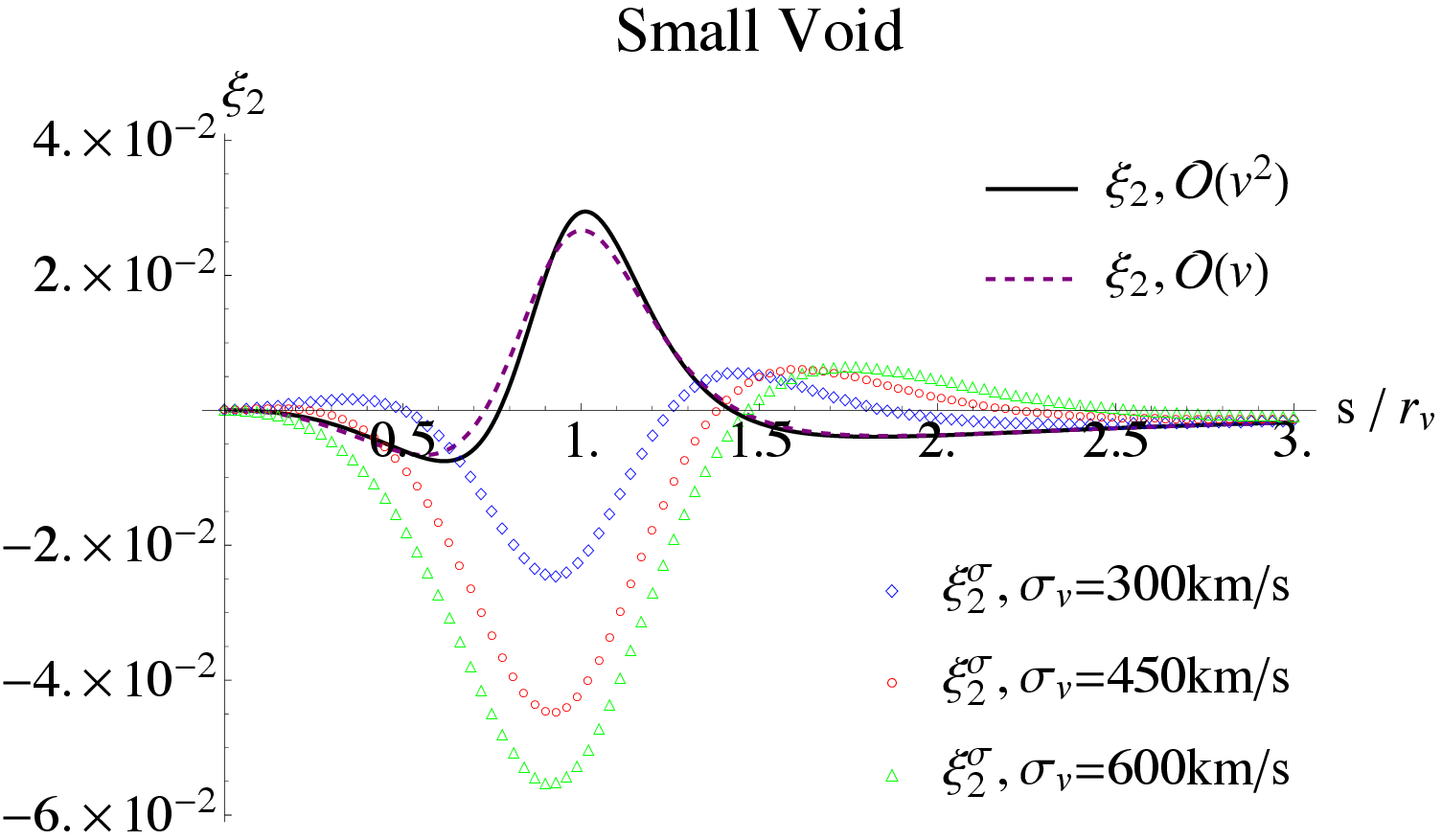}
  \end{center}
   \vspace{-0.cm}
 \end{minipage}
\caption{The plots of multipoles by applying void profile from Ref.\cite{UVP} under different circumstances
  on our formulation, where the parameters chosen are listed in Table II. The upper panels show the large
  size void, the middle panels are the medium size void, and the lower panels are the small size void.
  From the left panels to the right panels, we show the monopole, the dipole, and the quadrupole,
  respectively. The large size void is similar to the void model in Sec.~\ref{sec:profile_VIMOSP}.
\label{fig:xi_ell_hamaus}
}
\end{figure}

\begin{figure}
\begin{minipage}{0.3\hsize}
  \begin{center}
    \includegraphics[width=\textwidth,height=3.5cm]{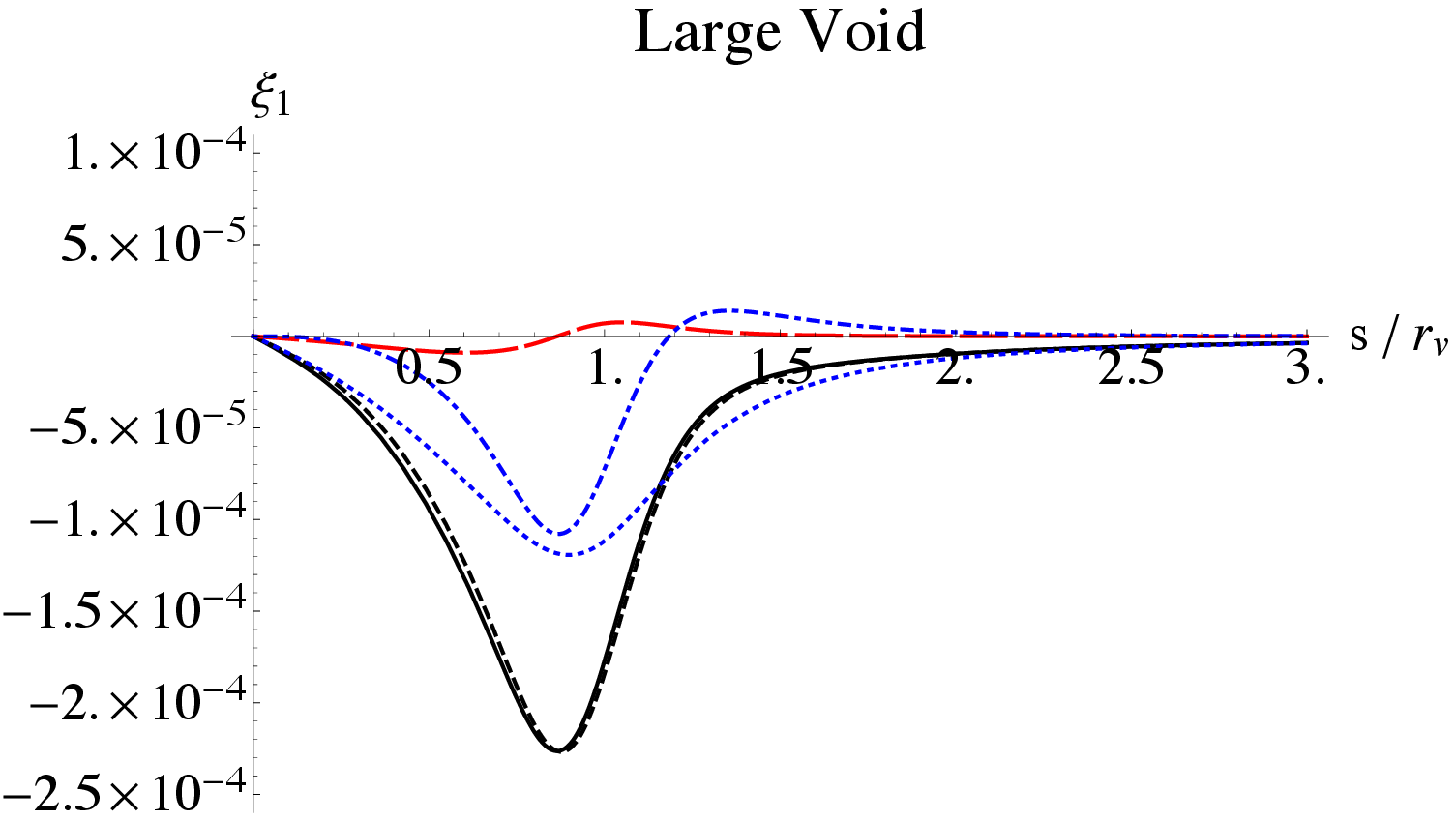}
  \end{center}
   \vspace{-0.cm}
 \end{minipage}
\hspace{0cm}
\begin{minipage}{0.3\hsize}
  \begin{center}
  \includegraphics[width=\textwidth,height=3.5cm]{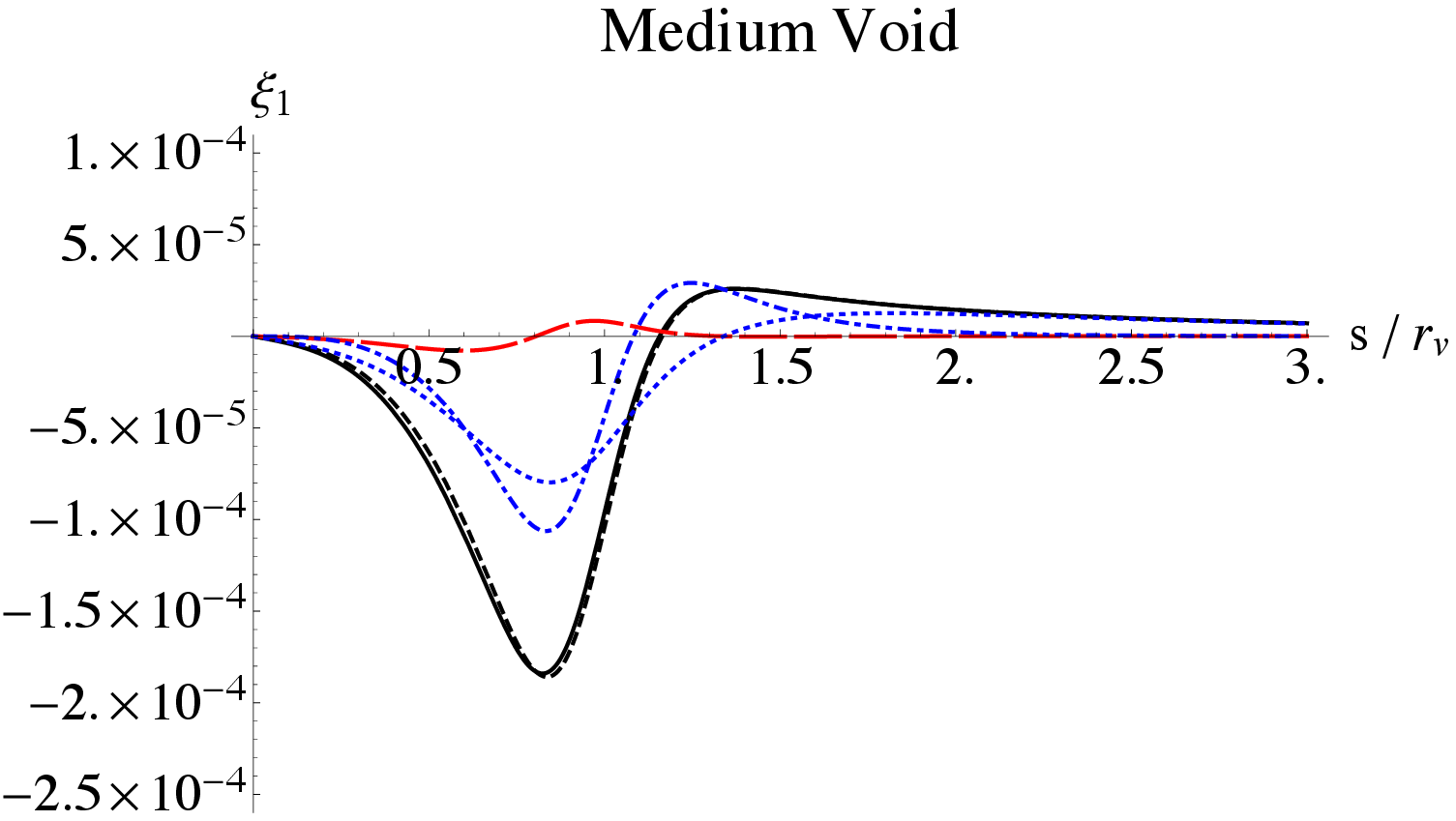}
  \end{center}
   \vspace{-0.cm}
 \end{minipage}
\hspace{0cm}
 \begin{minipage}{0.35\hsize}
  \begin{center}
  \includegraphics[width=\textwidth,height=3.5cm]{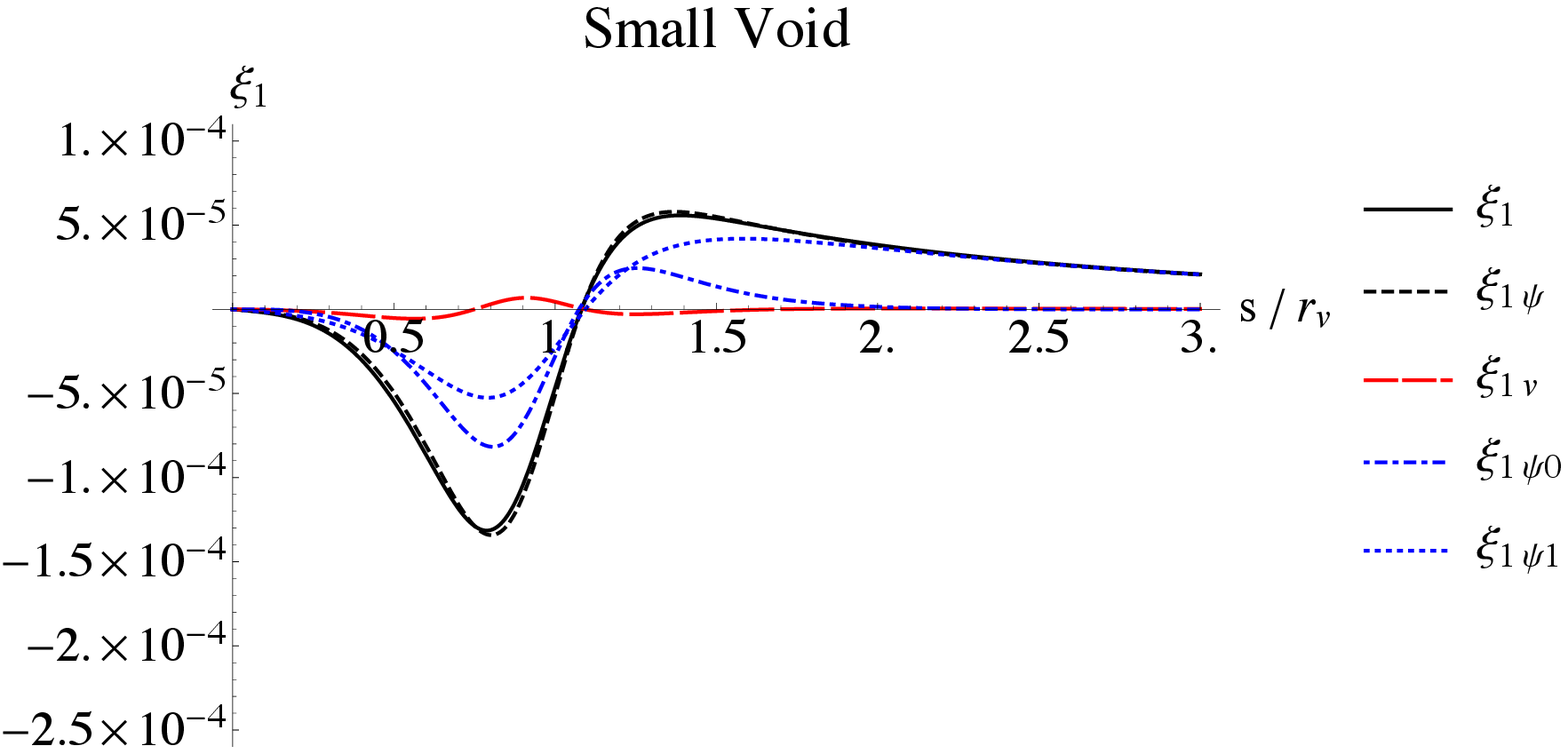}
  \end{center}
   \vspace{-0.cm}
 \end{minipage}
\caption{Details of the dipole components similar to Figure~\ref{fig:xi1_gvtot_g0g1},
where $\xi_{1}(=\xi_{1\psi}^{(s)}+\xi_{1v}^{(s)})$, $\xi_{1\psi}^{(s)}(=\xi_{1\psi0}^{(s)}+\xi_{1\psi1}^{(s)})$
is the dipole for the large size void, the medium size void, and the small size void with
Ref.~\cite{UVP}, from the left to right, respectively.
See Eqs.~(\ref{eq:xi1psi0psi1}) and (\ref{eq:xi1gv}) for the definition.
\label{fig:xi1_gvtot_g0g1_hamaus}
}
\end{figure}

Applying the void profiles to our formulation, we investigate the behaviors and the
properties of the multipoles defined by Eqs.~(\ref{eq:defmp}) and (\ref{eq:defmpsig})
in a similar way to what we have done in Sec.~\ref{sec:profile_VIMOSP}.
Figure~\ref{fig:xi_ell_hamaus} shows the results for the multipoles
and Figure~\ref{fig:xi1_gvtot_g0g1_hamaus} shows the details of comparison for the dipole
component in a similar way to Figures~\ref{fig:xi0sig}-\ref{fig:xi1sig}~and~
\ref{fig:xi1_gvtot_g0g1} in Sec.~\ref{sec:profile_VIMOSP}.
The profile of the model in the previous Sec.~\ref{sec:profile_VIMOSP}
is similar to the large size void, so their multipoles' behavior is almost
the same. The multipoles of the medium size void and small size void are similar
to the large size void; in particular, the dipole is dominated by the
gravitational redshift for all three void sizes.

\begin{figure}[t]
  \begin{center}
  \includegraphics[width=105mm]{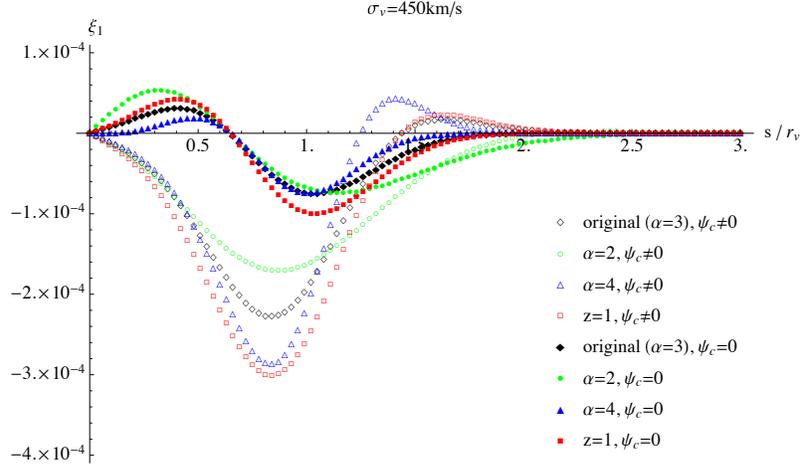}
  \end{center}
 \caption{
Dependence of the dipole $\xi_1^{\sigma}$ on different model
parameters for Eq.~(\ref{eq:profile_Hawken}) in Sec.~\ref{sec:profile_VIMOSP}
with fixed velocity dispersion $\sigma_v=450~{\rm km/s}$,
and different sets of model parameters $(\alpha=3,~z=0.5)$;
$(\alpha=2,~z=0.5)$; $(\alpha=4,~z=0.5)$; and $(\alpha=3,~z=1)$.
Both cases $\psi_c\neq0$ and $\psi_c=0$ are shown.
All the curves treat univariate parameter changes compared to the
original parameter set in Table I.
  \label{fig:xi1sig_param_psic}
 }
\end{figure}
\begin{figure}[h]
  \begin{center}
  \includegraphics[width=105mm]{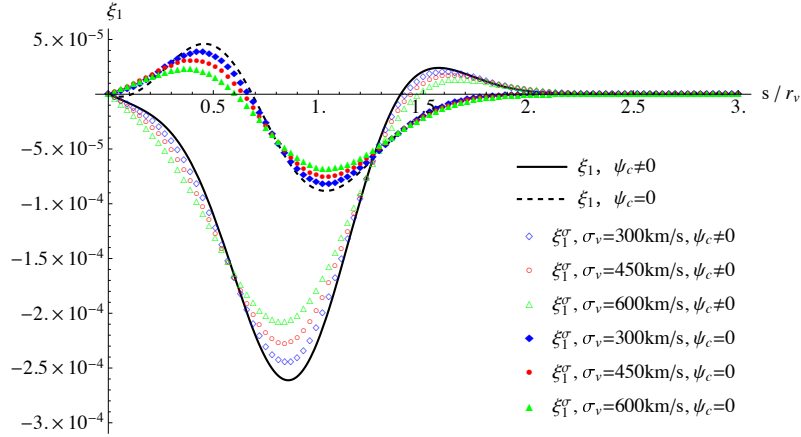}
  \end{center}
  \caption{
    Same as Figure~\ref{fig:xi1sig} but with the additional case under the
    condition $\psi_c=0$ for comparison.
  \label{fig:xi1sig_psic}
 }
\end{figure}
\begin{figure}[h]
  \begin{center}
  \includegraphics[width=105mm]{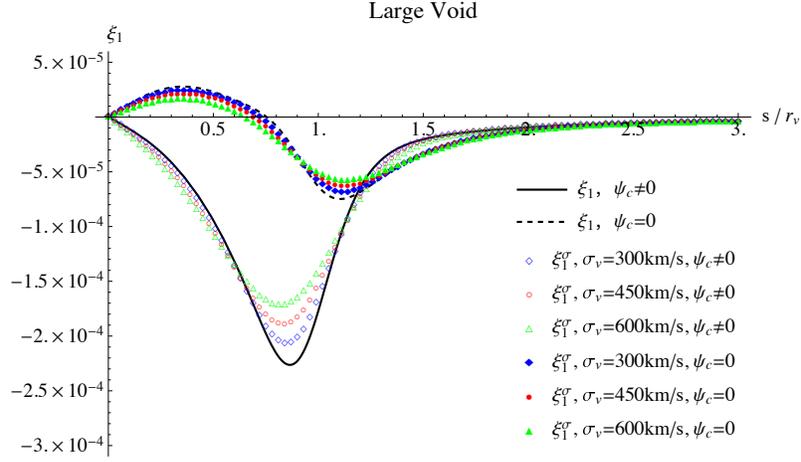}
  \end{center}
  \caption{
    Same as Figure~\ref{fig:xi1sig_psic} but for the large size void.
  \label{fig:xi1sig_psic_large}
 }
\end{figure}
\begin{figure}[t]
  \begin{center}
  \includegraphics[width=110mm]{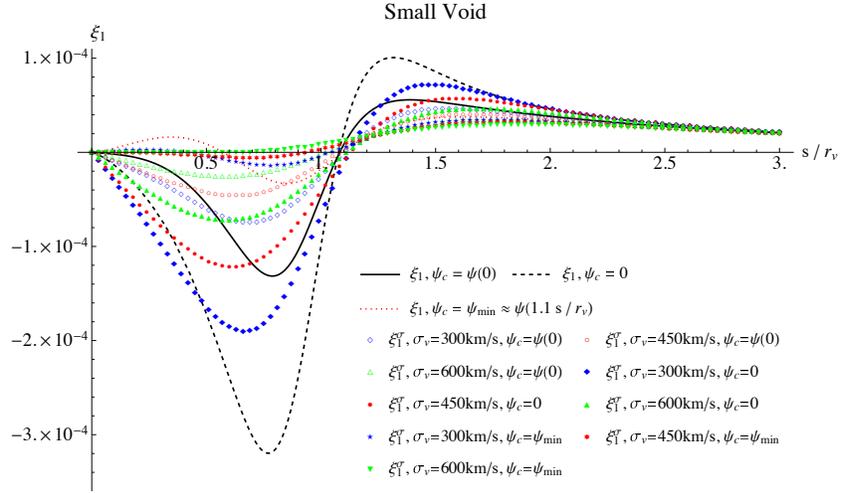}
  \end{center}
  \caption{
    Same as Figure~\ref{fig:xi1sig_psic} but for the small size void,
    with the additional case under the condition and
    $\psi_c=\psi_{\rm min}$ for comparison.
  \label{fig:xi1sig_psic_small}
 }
\end{figure}
\begin{figure}[t]
  \begin{center}
  \includegraphics[width=110mm]{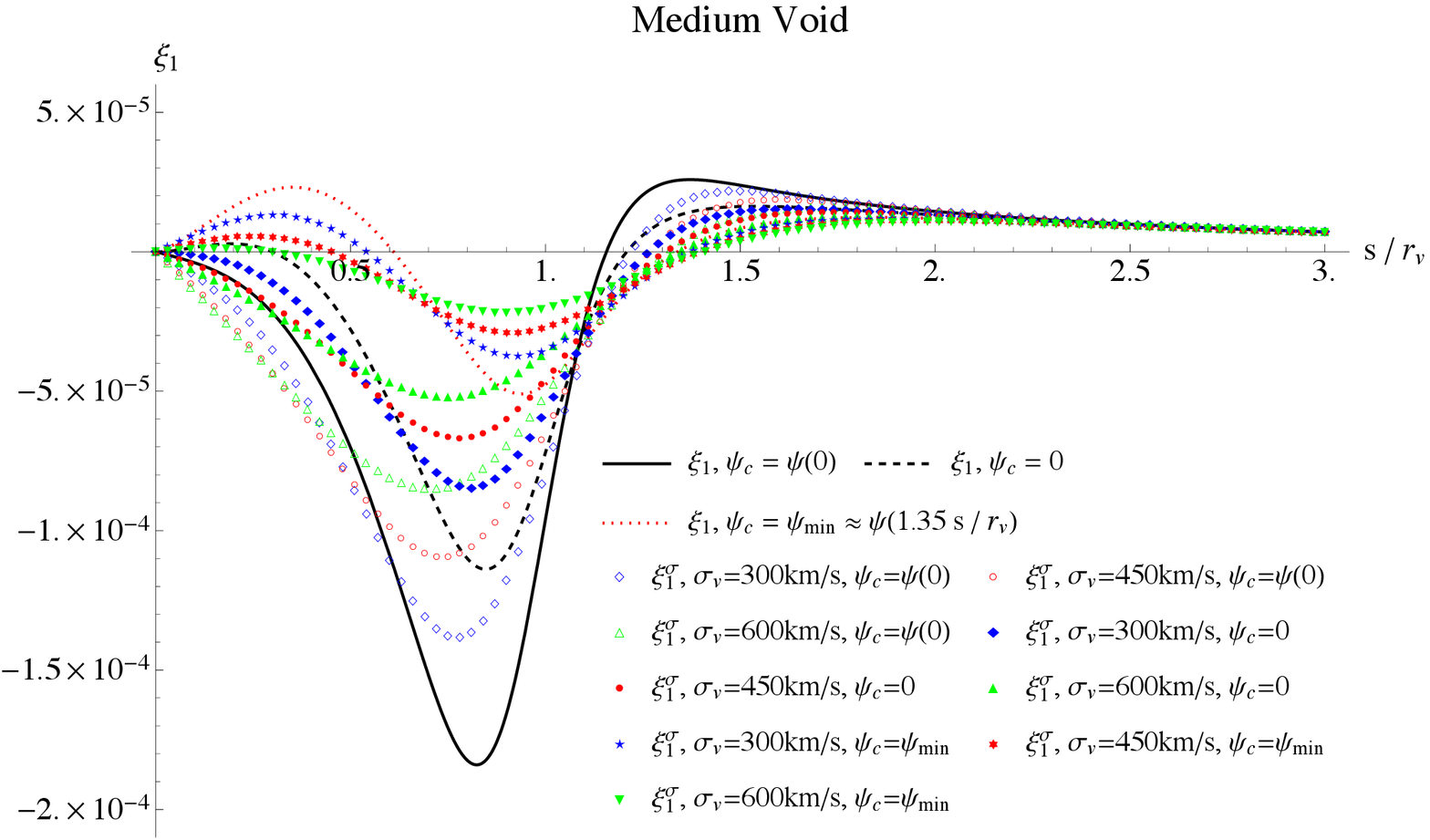}
  \end{center}
  \caption{
    Same as Figure~\ref{fig:xi1sig_psic} but for the medium size void,
    with the additional case under the condition $\psi_c=\psi_{\rm min}$
    for comparison.
  \label{fig:xi1sig_psic_medium}
 }
\end{figure}

\section{Discussions}
\label{sec:discu}
In the previous section, we demonstrated how the higher order terms of the
peculiar velocity and the gravitational potential influence the void-galaxy
correlation functions through redshift space distortions.
The higher order effect is not very significant for the monopole and the quadrupole
components. However, the most interesting finding from the higher order effects is that
the dipole component in the void-galaxy cross-correlation functions, dominantly
reflects the gravitational potential through the gravitational redshift,
as demonstrated in Figures~\ref{fig:xi1sig} \textemdash \ref{fig:xi1_gvtot_g0g1_hamaus}.

Photons from the central region of a large size void are blue-shifted compared with photons
from the mean density region. This effect of the gravitational redshift
is the origin of the dipole signal in the void-galaxy cross-correlation function.
Most importantly, the dipole signal is therefore determined by the gravitational
potential profile of the void.
However, the dipole signal is not simply dominated by the gravitational potential
$\psi(r)$.
The contribution of the gravitational redshift to the dipole,
Eq.~(\ref{eq:xi1_g}), is given by the combination of the
two terms,
\begin{eqnarray}
  \xi_1^{(s)}(s) \simeq \xi_{1\psi}^{(s)}(s)=\xi_{1\psi0}^{(s)}(s)+\xi_{1\psi1}^{(s)}(s)=
  \frac{\psi'(s)}{3\mathcal{H}}(1+\xi(s))+
  \frac{\psi(s)-\psi_c}{3\mathcal{H}}\xi'(s),
  \label{eq:xi1_approx}
\end{eqnarray}
namely, the terms from the gravitational potential $\psi(s)$ and
its gradient $\psi'(s)$ contribute equally to the dipole signal, as is demonstrated
in Figures~\ref{fig:xi1_gvtot_g0g1}~and~\ref{fig:xi1_gvtot_g0g1_hamaus} .

Although we restrict ourselves to the predictions
based on general relativity and discuss aspects of the
robustness of our theoretical predictions, measurements of
the gravitational potential provides a unique chance to
test general relativity and other gravity theories.
Moreover, these measurements may contribute to the clarification of bias on
calibration of cosmological parameters due to local gravitational environments
as mentioned previously.
Theories kept in the linear regime for low-density regions associated
with voids will make such tests less difficult compared with
those which require nonlinear modeling of clustering of galaxies.
To be a useful tool for testing gravity, the robustness of theoretical
predictions is necessary.
In Sec.~\ref{sec:profile_VIMOSP}, we fixed the model parameters of a void (see Table I).
Hence, it will be useful to check the robustness of our prediction of the results against
the dependence on the model parameters in Table I.


In Figure~\ref{fig:xi1sig_param_psic}, the dipole signal $\xi_1^\sigma$ is plotted
by fixing the velocity dispersion $\sigma_v=450~{\rm km/s}$, by varying
the set of parameters as $(\alpha=3, z=0.5)$; $(\alpha=2, z=0.5)$;
$(\alpha=4, z=0.5)$; and $(\alpha=3, z=1)$, based on the model of
Eq.~(\ref{eq:profile_Hawken}) in Sec.~III A.
We see that the dipole signal
changes with the different values of $\alpha$, which characterizes the
steepness of a void potential wall, and also the value of the redshift $z$,
which corresponds to different cosmological epoch.
As the void potential becomes steeper, the gravitational potential takes
a larger amplitude inside a void in the general $\psi_c\neq0$ case.
In the cold dark matter model with a
cosmological constant, the void gravitational potential decreases in
proportion to $D_1(a)/a$. These properties explain the
redshift dependence of the dipole in Figure~\ref{fig:xi1sig_param_psic}.
Nevertheless, our conclusions are not qualitatively altered by the
choice of these parameters. Furthermore, our conclusion is robust
against the change of model between Secs.~\ref{sec:profile_VIMOSP}~and~\ref{sec:profile_UVP}.

We next consider the importance of
the choice of the center of a void.
In Figures~\ref{fig:xi1sig} and \ref{fig:xi1_gvtot_g0g1}, we showed the case
in which the gravitational redshift of the center of a void $\psi_c$
is nonzero, i.e., the case $\psi_c=\psi(0)\neq0$.
This is the case when the shift of the center of a void through
the gravitational redshift is included as the extreme case.
However, this term depends on the strategy employed to define
the void center. For example, when the center of a void is
determined using galaxies far from the central region, the position of
the center is determined without the information $\psi_c$,
irrespective of the gravitational redshift of $\psi_c$.
In such a case, we should assume $\psi_c=0$.
Figure \ref{fig:xi1sig_psic} compares the case with/without the term
$\psi_c/{\cal H}(z_c)$.
These results indicate that the choice for the center of a void
is not trivial for the dipole signal.

The void density contrast profile of Eq.~(\ref{eq:profile_Hawken}) always
generates a gravitational potential with a positive sign.
However, the small size void and the medium size void of the best-fit profile
in Ref.~\cite{UVP} in Sec.~\ref{sec:profile_UVP}
predict the gravitational potential with a negative sign, $\psi(r)<0$,
as is demonstrated in Appendix~\ref{sec:appenc}.
This model allows us to test the case when the gravitational redshift from the
center of the small and medium size voids may be redshifted rather than blueshifted.
We further check the impact of the choice of the center of a void with the model
in Sec.~\ref{sec:profile_VIMOSP}, adopting the profile with the parameters
in Table II for testing the robustness of our prediction.
Figures~\ref{fig:xi1sig_psic_large}, \ref{fig:xi1sig_psic_small},
and \ref{fig:xi1sig_psic_medium}
show the impact of the choice of the void center on the dipole signal
for the model in Sec.~\ref{sec:profile_UVP}.
Figure~\ref{fig:xi1sig_psic_large} is the case of a large size void,
which corresponds to the model in Sec.~\ref{sec:profile_VIMOSP}, while
Figures \ref{fig:xi1sig_psic_small} and \ref{fig:xi1sig_psic_medium} are
show the small size void and the medium size void.

For the small size void and the medium size void, the void potential
possesses a global minimum at roughly $r \sim \mathcal{O}(1) r_v$ rather
than at infinity or at $r=0$($s=0$),
which is demonstrated in Figure~\ref{fig:velocity_potential} in Appendix~\ref{sec:appenc}.
In Figure~\ref{fig:xi1sig_psic_small},
we plot the dipole signal for the small size void
when the void center is sampled at the minimum of the gravitational
potential $\psi_c=\psi_{\rm min}$ at $s \approx 1.1 r_v$. In this case, the
dipole signal becomes small though it is nonvanishing.
In particular, the amplitude becomes even smaller when the random velocity is
included.
A more complicated situation occurs for the medium size void. The sign of the
gravitational potential changes from the negative to positive
in the interior region of the void (see Figure~\ref{fig:velocity_potential}
in Appendix~\ref{sec:appenc}).
In this case $\psi_c$ could be either positive or negative.
In Figure~\ref{fig:xi1sig_psic_medium},
we plot the dipole signal while adopting $\psi_c=\psi_{\rm min}$
in a similar way to Figure~\ref{fig:xi1sig_psic_small}.
Thus the dipole signal depends on the choice of $\psi_c$,
though it is nonvanishing.

The choice of the void center depends on the algorithm used in data analysis.
We need to investigate the behavior of the dipole signal by adopting the realistic
algorithm for finding the center of a void with mock catalogs of galaxies
from numerical simulations. Although this is beyond the scope of the present paper,
we briefly discuss this problem.
In a realistic analysis of void-finding algorithms, there are two
algorithms best suited to define the center of a void.
One is the lowest-density center, which is determined
as the center of the lowest density galaxy in a void and its three
most adjacent neighboring galaxies \cite{NadaHC,Nadathur:2016nqr}.
This algorithm is equivalent to finding the largest empty sphere that can be
found within the void. Another is the volume-weighted barycenter which is mainly
used for irregular-shaped voids (Ref.~\cite{Nadathur:2016nqr}).
Roughly, the first definition of the center by the minimum density center
corresponds to the case including $\psi_c$ as the maximum, while the second
volume-weighted barycenter might correspond to the case without $\psi_c$.
This is because $\psi_c$ should be included
when the center of a void is determined by the information about the central region,
while $\psi_c$ should not be included when the center is determined by the
information about the outskirt region of a void.

In a realistic situation, each void is not spherically symmetric, and statistical
errors should be included in identifying the center of the void as well as $\psi_c$.
To consider such an effect, we may be able to consider a probability
distribution function for it in a similar way to
Eq.~(\ref{eq:gausstreaming}) as
\begin{eqnarray}
  1+\xi^{\sigma,P}(s,\mu) = \int\left(\int\frac{1+\xi^{(s)}(s^{\sigma},\mu^{\sigma})}{\sqrt{2\pi}\sigma_v} \exp \left(
  -\frac{\vpara^2}{2\sigma_v^2} \right)\dif \vpara\right)P(\psi_c)\dif\psi_c,
  \label{eq:gausstreamingplus}
\end{eqnarray}
where $\xi^{\sigma,P}(s,\mu)$ stands for the void-galaxy cross-correlation with
consideration of the probability distribution on the $\psi_c$ term.
However, the probability distribution function of $\psi_c$, i.e.,
$P(\psi_c)$ may be highly model dependent and deviate
from the Gaussian distribution. So we would rather leave this part
to more sophisticated consideration in the future.

Finally in this section, we discuss the influence of spherical averaging
or stacking of the void profile on the dipole signal. In Refs.~\cite{Wojtak2016,Cautun},
the authors argued that the conventional way of spherical stacking of void
may lead to a stacked profile with steeper central density contrast
  and milder transition to the
high-density ridge at the void boundary compared with the actual situation.
In Ref.~\cite{Cautun}, the authors introduced a boundary profile closer to the
actual individual void obtained by stacking over distances of volume elements
from the void boundary rather than spherical averaging over its center,
which has a flat core with a sharp transition to the high-density ridge at the void
boundary. This is very similar to the profile of Eq.~(\ref{eq:profile_Hawken}) in
Sec.~\ref{sec:profile_VIMOSP} with a very large value of $\alpha$.

We find that the gravitational potential of a profile similar to Ref.~\cite{Cautun}
is flatter than that of the spherically stacked profile and almost zero
at the exterior regions of the void.
If the void center is determined with a profile like this by tracers at the exterior
regions of the void, for example, at $s \gtrsim r_v $,
$\psi_c$  will be
closer to zero than that with the spherically stacked profile.
This result suggests that the influence of the choice of void center
characterized by $\psi_c$ will be smaller if we sample the voids
using tracers from exterior regions by the method in Ref.~\cite{Cautun}.
Methods like this could make the dipole signal less sensitive to the choice of void center.
Thus, the strategy to choose the void center changes the dipole signal,
which causes some difficulty in the comparison of our results with observations,
although the monopole and the quadrupole do not depend on this choice.

\section{Summary and Conclusion}
\label{sec:concl}
In this paper, we have presented an analytic model for the void-galaxy cross-correlation
function in redshift space including the higher order terms of the peculiar
velocity and the gravitational potential through redshift space distortions.
By adopting specific models for a void density profile, including a universal best-fit profile
which can produce infall velocities in the linear regime of density perturbations, we have
quantitatively demonstrated the influence of the higher order terms on the multipole
components of the void-galaxy cross-correlation.
In particular, we have found that the dipole signal dominantly reflects the
gravitational potential through the gravitational redshift.
Our conclusion is qualitatively robust against the change of the model parameters and the void profiles.
However, we have also discussed the possible dependence of the dipole
signal on the algorithm for determining the center of a void.
This dependence should be investigated with the use of numerical simulations
with mock catalogs by adopting a practical algorithm to determine the center of a
void, including other systematic errors. The idea of consideration for the
probability distribution function of the $\psi_c$ term may be helpful, but this is left as a future investigation.
However, in principle, our finding presents the possibility of a new approach to direct
measurements of the gravitational potential of voids.

In Ref. \cite{Hamaus}, the monopole and the quadrupole multipoles of the
void-galaxy cross-correlation function were measured with the SDSS
III LOWZ sample and the CMASS sample. Our formulation may serve as an extension
to the analytic theory for quantitative analysis on these measurements with higher order accuracy.
The error bars of the multipoles are quite small; thus, we may detect the dipole
component in future analysis. Such an analysis would help in the understanding
of voids and provide a test for general relativity and cosmological models
in combination with observations using other methods, e.g.,
measurement of the thermal Sunyaev-Zel'dovich effect around voids
and the weak lensing measurements of voids \cite{Alonso}.

\vspace{2mm}
{\it Acknowledgments.}---
This work is supported by MEXT/JSPS KAKENHI Grants No. 15H05895, No. 17K05444, and
No. 17H06359 (K. Y.). We thank A. Taruya, S. Saito, and D. Parkinson for
useful communications during the workshop YITP-T-17-03.
We thank an anonymous referee for helpful comments, which improved the manuscript.


\appendix

\newtagform{WithA}{(}{)}
\usetagform{WithA}
\newcommand{\Aref}[1]{(A\ref{#1})}

\section{Complementary Derivations for Eq.~(\ref{eq:xissmu})}
\label{sec:appena}

We will present some details for derivations in Sec.~\ref{sec:formu}.
Since we adopted the plane parallel approximation, we will assign the
three-dimensional coordinates $\vec S$ and $\vec R$
in redshift space and real space, respectively, as
\begin{eqnarray}
  \vec S=S\vec \gamma + \vec x_\perp,
  \nonumber\\
  \vec R=\chi \vec \gamma + \vec x_\perp,
\end{eqnarray}
where $\vec x_\perp$ is the coordinate perpendicular to the line-of-sight direction.
Denoting the position of the center of a void as $\vec R_c=R_c \vec \gamma$ and
$\vec S_c=S_c \vec\gamma$ in real space and the redshift space, respectively
we have
\begin{eqnarray}
  \vec s+\vec S_c=\vec r+\vec R_c
  +\biggl[{(1+z)\over H(z)}\left(\bm \gamma\cdot {\bm v}+{1\over 2}\bm v^2+(\bm \gamma\cdot {\bm v})^2
  -\psi\right)
  -{H'(z)\over 2H^2(z)}(1+z)^2(\bm \gamma\cdot {\bm v})^2\biggr]\vec \gamma,
\end{eqnarray}
from Eq.~(\ref{eq:bigs}), where we adopted the coordinate system with
its origin at the center of a void $\vec r$ and $\vec s$ in
the real space and the redshift space, respectively.

When we choose $S_c=R_c=\chi_c=\int_0^{z_c}dz'/H(z')$,
we have
\begin{eqnarray}
  \vec s=\vec r+\biggl[{(1+z)\over H(z)}\left(\bm \gamma\cdot {\bm v}+{1\over 2}\bm v^2+(\bm \gamma\cdot {\bm v})^2
  -\psi\right)
  -{H'(z)\over 2H^2(z)}(1+z)^2(\bm \gamma\cdot {\bm v})^2\biggr]\vec \gamma.
\end{eqnarray}
Introducing the conformal Hubble parameter ${\cal H}=aH$
and using the relation $a=1/(1+z)$ will slightly simplify the verbose expression as
Eq.~(\ref{eq:coorred0}) that can be expressed as
\begin{eqnarray}
  \vec s=\vec r+\biggl[
    {\bm \gamma\cdot {\bm v}\over {\cal H}(z)}
    +{1\over 2}{\bm v^2\over {\cal H}(z)}+{(\bm \gamma\cdot {\bm v})^2
      \over {\cal H}(z)}
  -{\psi\over {\cal H}(z)}
  -{H'(z)\over 2{{\cal H}^2(z)}}(\bm \gamma\cdot {\bm v})^2\biggr]\vec \gamma.
\label{eq:coorred0ap}
\end{eqnarray}
This is for the case of the center of a void and the origin of the coordinate
does not change between the real space and the redshift space.

When we choose $S_c+\psi_c/{{\cal H}(z_c)}=R_c=\chi_c=\int_0^{z_c}dz'/H(z')$, we have
\begin{eqnarray}
  \vec s=\vec r+\biggl[
    {\bm \gamma\cdot {\bm v}\over {\cal H}(z)}
    +{1\over 2}{\bm v^2\over {\cal H}(z)}+{(\bm \gamma\cdot {\bm v})^2
      \over {\cal H}(z)}
  -{\psi\over {\cal H}(z)}
  -{H'(z)\over 2{{\cal H}^2(z)}}(\bm \gamma\cdot {\bm v})^2+{\psi_c \over {\cal H}(z_c)}\biggr]\vec \gamma,
  \label{eq:coorredap}
\end{eqnarray}
This is for the case when the center of the void and the origin of the coordinate
shift from the real space to the redshift space.

Note that Eq.~(\ref{eq:coorred0ap}) is reproduced by setting
$\psi_c=0$ in Eq.~(\ref{eq:coorredap}). Then we present
the formulation with Eq.~(\ref{eq:coorredap}) in the following section.

Then the component parallel to $\vec \gamma$ in Eq.~(\ref{eq:coorred}) is in fact
\begin{eqnarray}
  \spara=\rpara+\biggl[
    {\bm \gamma\cdot {\bm v}\over {\cal H}(z)}
    +{1\over 2}{\bm v^2\over {\cal H}(z)}+{(\bm \gamma\cdot {\bm v})^2
      \over {\cal H}(z)}
  -{\psi\over {\cal H}(z)}
  -{H'(z)\over 2{{\cal H}^2(z)}}(\bm \gamma\cdot {\bm v})^2+{\psi_c \over {\cal H}(z_c)}\biggr].
\end{eqnarray}
Comparing this with the first line of Eq.~(\ref{eq:sdecomp2}), where $\spara=\rpara+\delta \rpara$,
we get Eq.~(\ref{eq:drpara}) for $\delta \rpara$.

Substituting the dimensionless quantity that we defined in Eq.~(\ref{eq:Vdless}) into Eq.~(\ref{eq:drpara}),
we write Eq.~(\ref{eq:srp2}) as
\begin{eqnarray}
\delta \rpara=    \tilde V(z_c,r)\rpara
  +{1\over 2}{\cal H}(z_c)\tilde V^2(z_c,r)r^2
  +{\cal H}(z_c)\tilde V^2(z_c,r)\rpara^2
  -{H'(z_c)\over 2}\tilde V^2(z_c,r)\rpara^2+{\psi_c \over {\cal H}(z_c)}
    -{\psi(r)\over {\cal H}(z_c)}.
  \nonumber
\end{eqnarray}
In $\delta \rpara$ in the previous derivation, all terms are $\mathcal{O}(v^2)$ terms except $\tilde V(z_c,r)\rpara \sim \mathcal{O}(v)$;
thus, we can write
\begin{eqnarray}
\delta \rpara = \tilde V(z_c,r)\rpara +\mathcal{O}(v^2),
  \label{eq:order_drpara}
\end{eqnarray}
which will be convenient in later derivations.

As an example of how we keep terms up to the order of $\mathcal{O}(v^2)$, we now consider the $\tilde V(z_c,r)\rpara$ term which frequently appears in the expression for $\delta\rpara$ as Eq.~(\ref{eq:srp2}).
Remembering Eq.~(\ref{eq:order_drpara}) will be very helpful in the process of keeping terms up to the order of $\mathcal{O}(v^2) \sim \mathcal{O}(\tilde V^2)$ in the following section.

Using Eq.~(\ref{eq:r}) for $r$ and Eq.~(\ref{eq:rpara}) for $\rpara$ repeatedly, together with Eq.~(\ref{eq:order_drpara}), we write
\begin{eqnarray}
  \tilde V(z_c,r)\rpara && =\tilde V(z_c,s-\mu\delta\rpara)(\spara-\delta\rpara)
  \nonumber \\
  &&\simeq \left(\tilde V(z_c,s)-\mu\delta\rpara \tilde V'(z_c,s)\right)(\spara-\tilde V(z_c,r)\rpara + \mathcal{O}(v^2))
  \nonumber \\
  &&\simeq \tilde V(z_c,s)\spara - \tilde V(z_c,s)\tilde V(z_c,r)\rpara - \mu \spara \tilde V'(z_c,s)\delta\rpara
  \nonumber \\
  &&\simeq \tilde V(z_c,s)\spara - \tilde V(z_c,s)\tilde V(z_c,s-\mu\delta\rpara)(\spara-\delta\rpara) - \mu \spara \tilde V'(z_c,s)\tilde V(z_c,r)\rpara.
  \label{eq:vrrp0}
\end{eqnarray}
In the above expression, noticing Eq.~(\ref{eq:order_drpara}) again, we further write for the second term in the expression above
\begin{eqnarray}
   \tilde V(z_c,s)\tilde V(z_c,s-\mu\delta\rpara)(\spara-\delta\rpara) &&\simeq \tilde V(z_c,s)\tilde V(z_c,s - \mu\delta\rpara)(\spara-\tilde V(z_c,r)\rpara )
   \nonumber \\
   && \simeq  \tilde V(z_c,s)\left(\tilde V(z_c,s) -\mu\delta\rpara \tilde V'(z_c,s)\right)\spara + \mathcal{O}(v^3)
   \nonumber \\
   && \simeq \tilde V(z_c,s)^2 \spara.
   \label{eq:part2vrrp}
\end{eqnarray}
Also, for the third term in Eq.~(\ref{eq:vrrp0}), using the definition for $\mu$ in Eq.~(\ref{eq:sdecomp1}),
transforming $\tilde V(z_c,r)\rpara$ similarly to Eq.~(\ref{eq:vrrp0}) will lead to
\begin{eqnarray}
  \mu \spara \tilde V'(z_c,s)\tilde V(z_c,r)\rpara && \simeq \frac{\spara^2}{s}\tilde V'(z_c,s)\tilde V(z_c,s)\spara + \mathcal{O}(v^3)
  \nonumber \\
  && \simeq \frac{\spara^3}{s}\tilde V'(z_c,s)\tilde V(z_c,s).
   \label{eq:part3vrrp}
\end{eqnarray}
Inserting Eqs.~(\ref{eq:part2vrrp}) and (\ref{eq:part3vrrp}) back into Eq.~(\ref{eq:vrrp0}) will lead us to Eq.~(\ref{eq:Vzr}).

With $r$ in Eq.~(\ref{eq:r})and $\tilde V(z_c,r)\rpara$ in Eq.~(\ref{eq:Vzr}) calculated up to the order of $\mathcal{O}(v^2)$,
we can now rewrite $\delta\rpara$ following the same previous procedure as Eq.~(\ref{eq:srp2}), which is a function of the redshift-space quantities $s$ and $\spara$ as Eq.~(\ref{eq:rsp}). Notice that $\psi \simeq \mathcal{O}(v^2)$, hence we have $\psi(r) \simeq \psi(s) $ directly.

Eq.~(\ref{eq:rpara}) and Eq.~(\ref{eq:rsp}) give the complete relation between $\rpara$ and $\spara$, so we can write
\begin{eqnarray}
  \rpara=&&\spara-\delta\rpara
  \nonumber\\
  =&&\spara-\Big[{\psi_c\over {\cal H}(z_c)}
    +\tilde V(z_c,s)\spara-\tilde V(z_c,s)^2 \spara-(\tilde V(z_c,s)\tilde V'(z_c,s)/s)
    \spara^3
    \nonumber\\
    &&+{1\over 2}{\cal H}(z_c)\tilde V^2(z_c,s)s^2
    +{\cal H}(z_c)\tilde V^2(z_c,s)\spara^2
    -{\psi(s)\over {\cal H}(z_c)}
    -{H'(z_c)\over 2}\tilde V^2(z_c,s)\spara^2\Big].
    \label{eq:rparaappen}
\end{eqnarray}
It is also obvious that
\begin{eqnarray}
&&  \spara = \mu s,
  \nonumber\\
&&  {\partial s \over \partial \spara}={\partial\sqrt{\spara^2+\sperp^2} \over \partial \spara}={1 \over 2 \sqrt{\spara^2+\sperp^2}}\cdot 2\spara={\spara \over s}=\mu.
\label{eq:sspara}
\end{eqnarray}
Combining Eqs.~(\ref{eq:rparaappen}) and~(\ref{eq:sspara}) we obtain Eq.~(\ref{eq:rstrans}).

On the other hand, relation given as
\begin{eqnarray}
  r&=&s-\mu\biggl({\psi_c\over {\cal H}(z_c)} + \tilde V(z_c,s)\spara- \tilde V^2(z_c,s)\spara
  -{\tilde V(z_c,s)\tilde V'(z_c,s)\over s}\spara^3
  +{1\over 2}{\cal H}(z_c)\tilde V^2(z_c,s)s^2
  \nonumber\\
  &&
  +{\cal H}(z_c)\tilde V^2(z_c,s)\spara^2
  -{\psi(s)\over {\cal H}(z_c)}
  -{H'(z_c)\over 2}\tilde V^2(z_c,s)\spara^2\biggr)+{\tilde V^2(z_c,s)\over 2s}\spara^2(1-\mu^2)
\end{eqnarray}
simply leads to Eq.~(\ref{eq:xir}).

\section{Potentially Useful Quantities}
\label{sec:appenb}
From Eq.~(\ref{eq:xissmu}), it might also be noted that for $\mu=0$, which stands for the direction perpendicular
to the line-of-sight direction, we have
\begin{eqnarray}
  \xi^{(s)}_{\sperp}&&=-\tilde V +\tilde V^2 + (1-\tilde V +\tilde V^2) \xi(s)
\nonumber\\
&&=(1+\xi(s))(1-\tilde V +\tilde V^2)-1.
\end{eqnarray}

The cosine angle between the galaxy displacement vector $\vec r$ from the void center and the line-of-sight $\vec \gamma$
in real space is
\begin{eqnarray}
  \mu_r \equiv {\rpara\over r}&=&{\mu s -\delta \rpara \over s-\mu\delta \rpara+\delta \rpara^2
  (1-\mu^2)/2s}
  \nonumber\\
  &\simeq&\mu-{\delta \rpara\over s}(1-\mu^2)+{3\over 2}\left({\delta \rpara\over s}
  \right)^2 \mu (1-\mu^2).
\end{eqnarray}


\section{Void Profiles Constructed from Ref.\cite{UVP}}
\label{sec:appenc}
\begin{figure}[ht]
\begin{minipage}{0.3\hsize}
  \begin{center}
  \includegraphics[width=\textwidth,height=3.7cm]{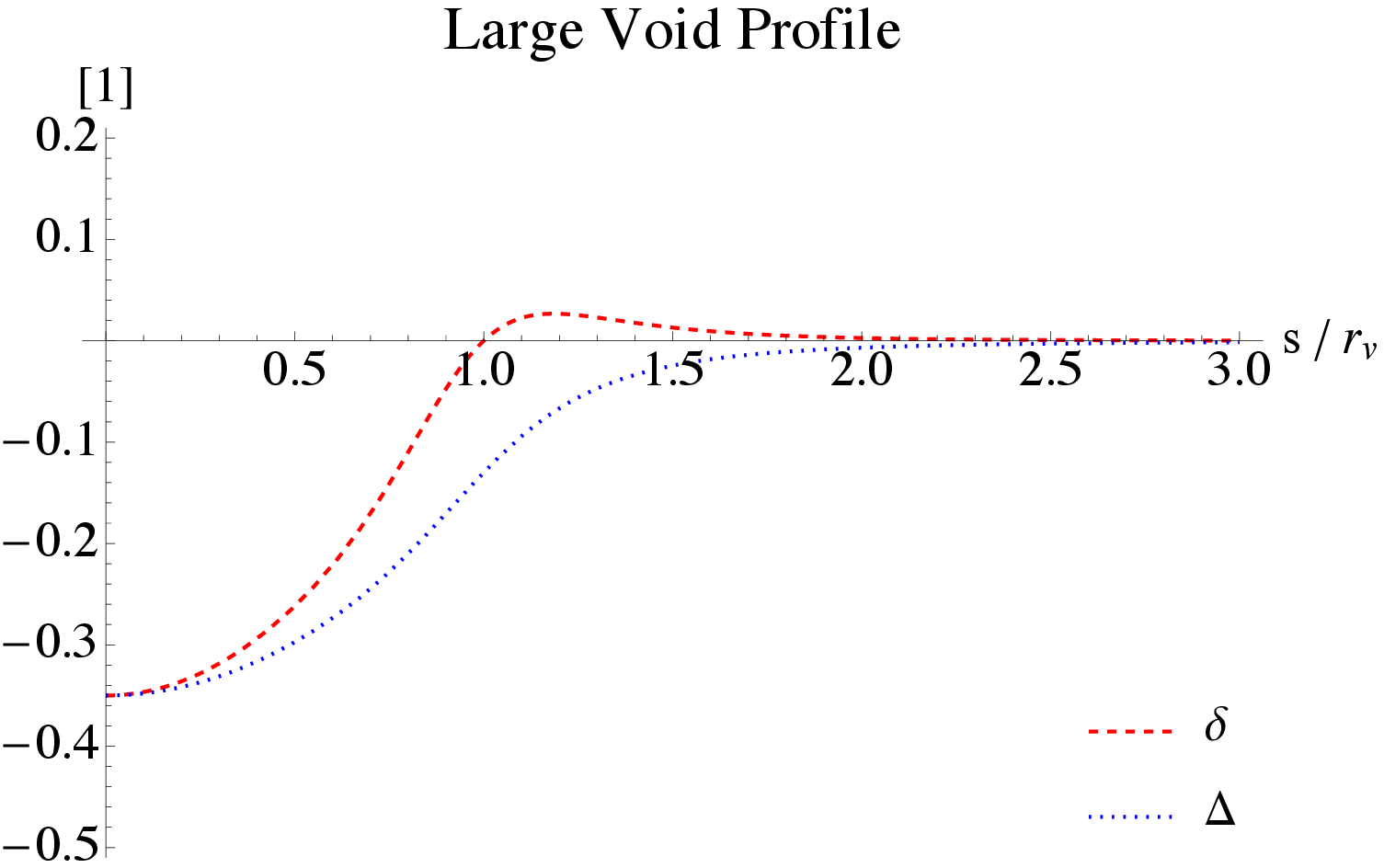}
  \end{center}
   \vspace{-0.cm}
 \end{minipage}
\hspace{0cm}
\begin{minipage}{0.3\hsize}
  \begin{center}
  \includegraphics[width=\textwidth,height=3.7cm]{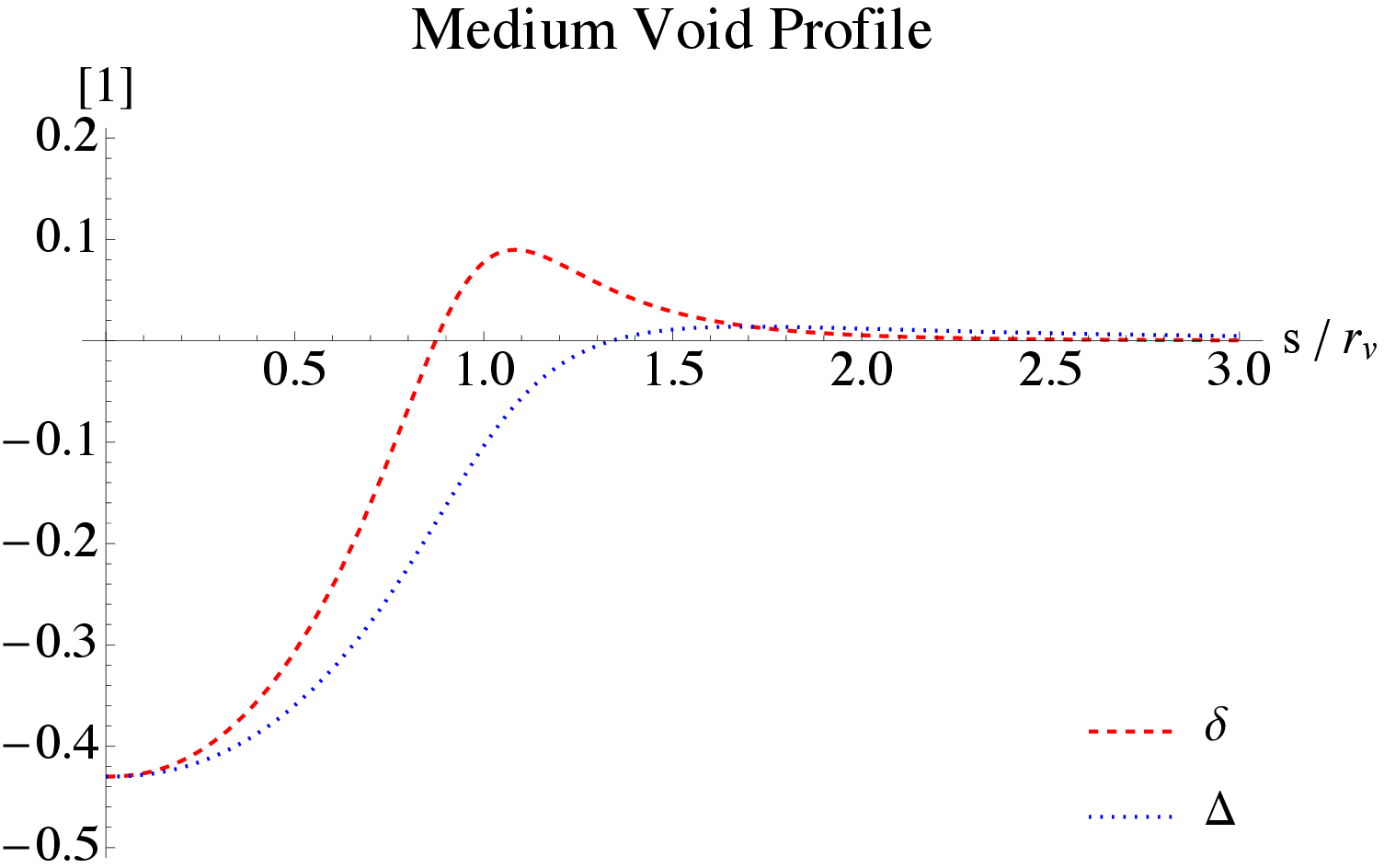}
  \end{center}
   \vspace{-0.cm}
 \end{minipage}
\hspace{0.2cm}
 \begin{minipage}{0.3\hsize}
  \begin{center}
  \includegraphics[width=\textwidth,height=3.7cm]{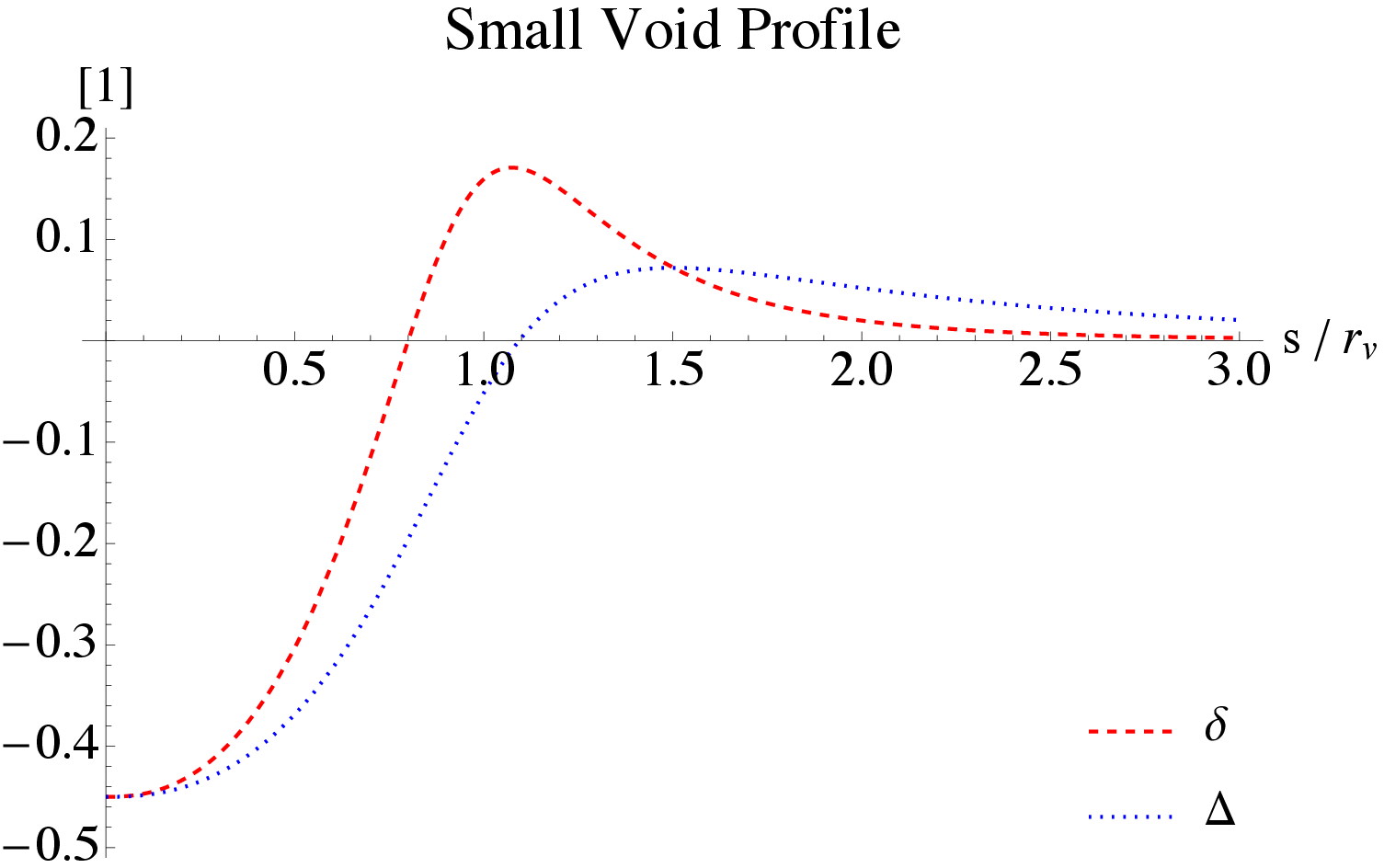}
  \end{center}
   \vspace{-0.cm}
 \end{minipage}
 \caption{The density contrast profiles as a function of $s/r_v$ for voids of different size
  with the parameters in Table II.
  The left, middle and right panels show the large size void, the medium size void,
  and the small size void, respectively.
\label{fig:profile_hamaus}
}
\end{figure}

\begin{figure}[ht]
\begin{minipage}{0.45\hsize}
  \begin{center}
  \includegraphics[width=\textwidth]{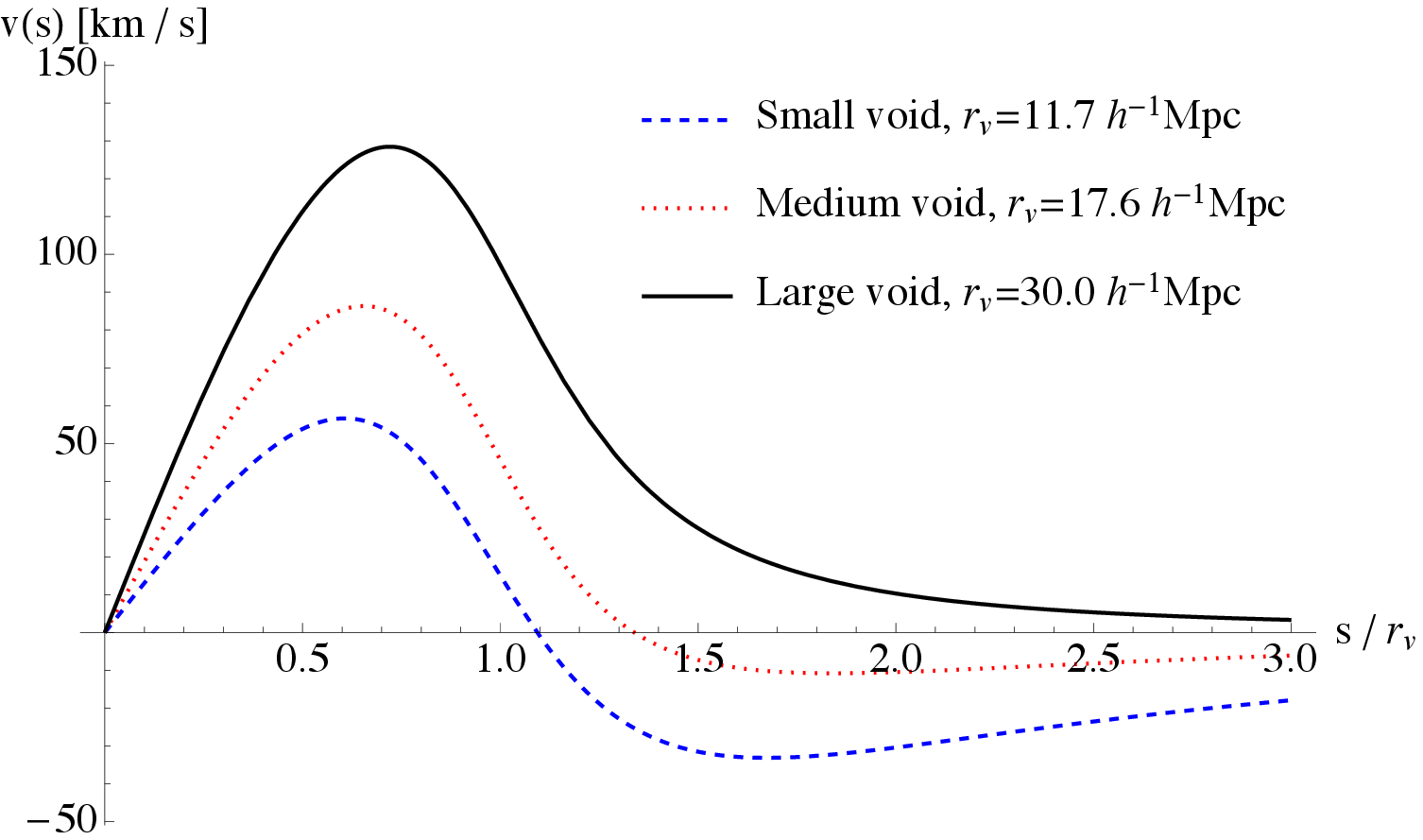}
  \end{center}
   \vspace{-0.cm}
\end{minipage}
\begin{minipage}{0.45\hsize}
    \begin{center}
    \includegraphics[width=\textwidth]{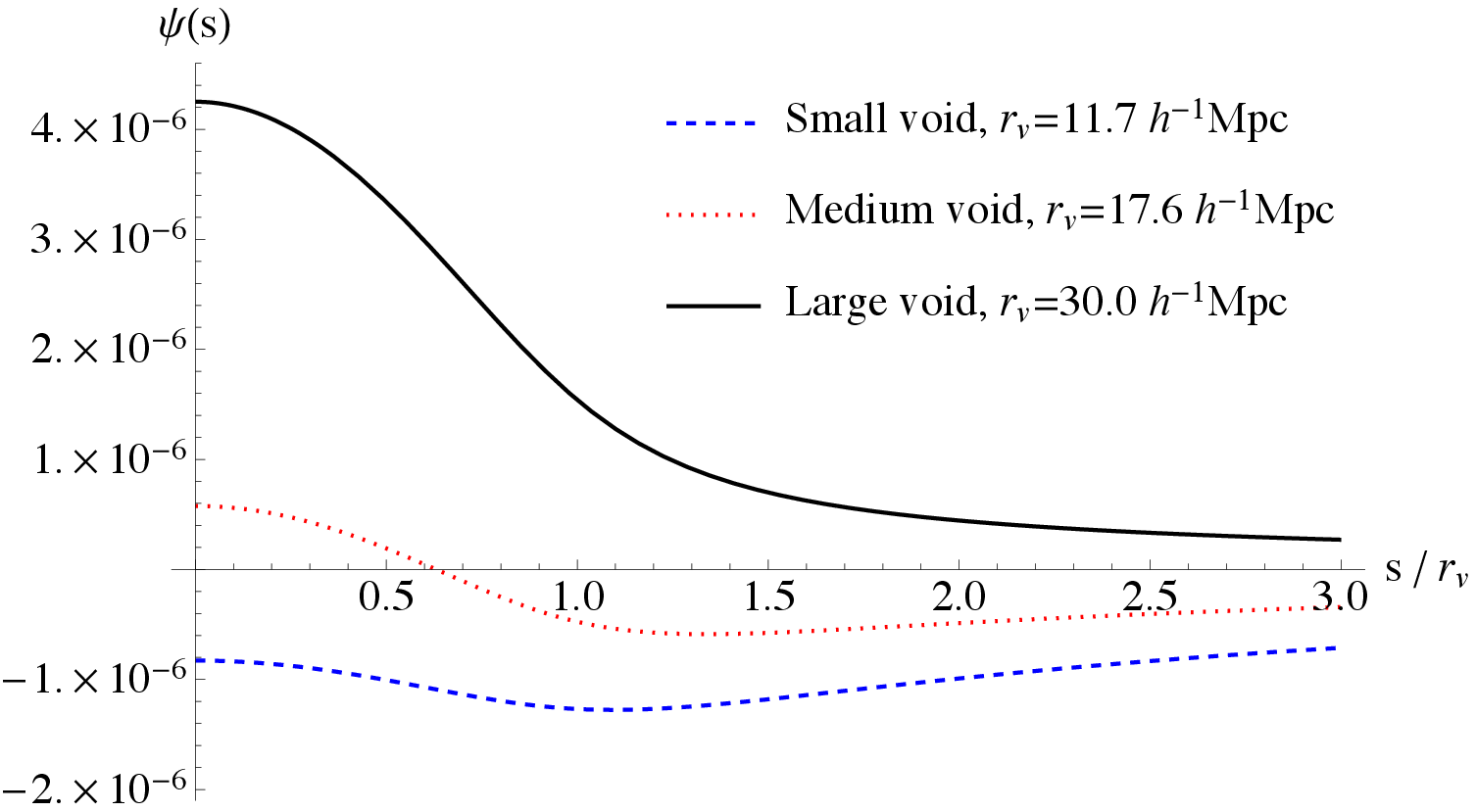}
    \end{center}
     \vspace{-0.cm}
\end{minipage}
\caption{
  The velocity profile (left panel), and the gravitational potential profile (right panel) as a function of
  $s/r_v$, for the small size void, the medium size void, and the large size void.
  \label{fig:velocity_potential}
 }
\end{figure}

We present the density profiles (Figure~\ref{fig:profile_hamaus})
and the velocity profile and the gravitational potential (Figure~\ref{fig:velocity_potential})
constructed from the model in Ref.~\cite{UVP}. Here we adopt the three typical sizes of voids,
the large size void, the medium size void, and the small size void, whose parameters are
listed in Table II. These three models correspond to those adopted in the text of the
present paper.

\vspace{8mm}

\end{document}